\documentclass[a4paper,10pt]{article}

\usepackage{epsfig}
\usepackage{epstopdf}
\usepackage{subfigure}
\usepackage{calc}
\usepackage{amssymb}
\usepackage{amstext}
\usepackage{amsmath}
\usepackage{amsthm}
\usepackage{multicol}
\usepackage{pslatex}
\usepackage[small]{caption}
\usepackage{color}
\usepackage[nohead, top=2.2cm, bottom=2.6cm, left=2.5cm, right=2.5cm]{geometry}
\usepackage{letterspace}
\usepackage{graphicx}
\usepackage{psfrag}
\usepackage{parcolumns}

\usepackage{enumerate}
\usepackage{enumitem}

\def\q5uad{\quad\quad\quad\quad\quad}

\newcommand*{\Mytab}{\hspace*{0.35cm}}

\title{WaterRPG: A Graph-based Dynamic Model \\ for Software Watermarking\vspace{0.4cm}}
\title{WaterRPG -- A Graph-based Dynamic Watermarking Model \\ for Software Protection -- Implementation and Evaluation\vspace{0.4cm}}
\title{A Graph-based Dynamic Watermarking Model for Software Protection using Function Calls  \vspace{0.4cm}}
\title{WaterRPG: a Graph-based Dynamic Watermarking \\ Model for Software Protection  \vspace{0.4cm}}
\title{WaterRPG: A Graph-based Dynamic Watermarking Model \\ for Software Protection  \vspace{0.4cm}}
\title{Protecting Software using the Graph-based Dynamic Watermarking Model WaterRPG    \vspace{0.4cm}}
\title{Protecting Software using the Graph-based Dynamic \\ Watermarking Model WaterRPG    \vspace{0.4cm}}
\title{The Graph-based Software Watermarking Model WaterRPG: Implementation and Evaluation   \vspace{0.4cm}}
\title{WaterRPG: A Graph-based Dynamic Model \\ for Software Watermarking  \vspace{0.4cm}}
\title{WaterRPG: A Graph-based Dynamic Watermarking \\ Model for Software Protection \vspace{0.4cm}}

\author{{\large Ioannis~Chionis \ \ \  Maria~Chroni \ \ \  Stavros~D.~Nikolopoulos}\\
{\normalsize \textit{Department of Computer Science and Engineering, University of Ioannina}}\\
{\normalsize \textit{P.O.Box 1186, GR-45110 \, Ioannina, Greece}}\\
{\normalsize \{ichionis, mchroni, stavros\}@cs.uoi.gr}
}

\date{}

\begin{document}

\maketitle

\begin{abstract}
\noindent Software watermarking involves embedding a unique identifier or, equivalently, a watermark value within a software to prove owner's authenticity and thus to prevent or discourage copyright infringement. Towards the embedding process, several graph theoretic watermarking algorithmic techniques encode the watermark values as graph structures and embed them in application programs. Recently, we presented an efficient codec system for encoding a watermark number $w$ as a reducible permutation graph $F[\pi^*]$ through the use of self-inverting permutations $\pi^*$. In this paper, we propose a dynamic watermarking model, which we call WaterRPG, for embedding the watermark graph $F[\pi^*]$ into an application program $P$. The main idea behind the proposed watermarking model is a systematic use of appropriate calls of specific functions of the program $P$. More precisely, for a specific input $I_{key}$ of the program $P$, our model takes the dynamic call-graph $G(P, I_{key})$ of $P$ and the watermark graph $F[\pi^*]$, and produces the watermarked program $P^*$ having the following key property: its dynamic call-graph $G(P^*, I_{key})$ is isomorphic to the watermark graph $F[\pi^*]$. Within this idea the program $P^*$ is produced by only altering appropriate calls of specific functions of the input application program $P$. We have implemented our watermarking model WaterRPG in real application programs and evaluated its functionality under various and broadly used watermarking assessment criteria.
The evaluation results show that our model efficiently watermarks Java application programs with respect to several watermarking metrics like data-rate, bytecode instructions overhead, resiliency, time and space efficiency. Moreover, the embedded watermarks withstand several software obfuscation and optimization attacks.

\end{abstract}

\section{Introduction}
\label{sec:introduction}

\noindent The rapid growth of World Wide Web users, the ease of distributing fast and in the original form digital content through internet, as well as the lack of technical measures to assure the intellectual property right of owners, has led to an increment in copyright infringement. Digital watermarking is a technique for protecting the intellectual property of any digital content, i.e., image, audio, video, software, text, ect. The main idea of digital watermarking is the embedding of a unique identifier into the digital content through the introduction of errors not detectable by human perception \cite{Book-CN10,CKLS96}. Although digital watermarking has made considerable progress and become a popular technique for copyright protection of multimedia information \cite{CKLS96}, research on software watermarking has recently received sufficient attention.

\vspace*{0.1in}
\noindent {\bf Software Watermarking.} Software watermarking is a technique that is currently being studied to prevent or discourage software piracy and copyright infringement. The software watermarking problem can be described as the problem of embedding a structure $w$ into a program $P$ and, thus, producing a new program $P_w$, such that $w$ can be reliably located and extracted from $P_w$ even after $P_w$ has been subjected to code transformations such as translation, optimization and obfuscation \cite{MC06}. More precisely, given a program $P$, a watermark $w$, and a key $k$, the software watermarking problem can be formally described by the following two functions: ${\tt embed}(P, w, k)$ $\rightarrow$ $P_w$ and ${\tt extract}(P_w, k)$ $\rightarrow$ $w$.

There are two main categories of software watermarking techniques namely {\it static} and {\it dynamic} \cite{CT99}. A static watermark $w$ is embedded inside program code in a certain format and it does not change during the program execution. On the other hand, a dynamic watermark $w$ is encoded in a data structure built at runtime (i.e., during program execution), perhaps only after receiving a particular input $I_{key}$; it might be retrieved from the watermarked program $P_w$ by analyzing the data structures built when program $P_w$ is running on input $I_{key}$. Moreover, depending on the behavioral properties of the embedded watermark $w$, software watermarking techniques can also be divided into other categories namely {\it robust} and {\it fragile}, {\it visible} and {\it invisible}, {\it blind} and {\it informed}, {\it focus} and {\it spread spectrum}; further discussion on the above software watermarking classification issues can be found in \cite{DM96,MC96,VVS01}.

\vspace*{0.1in}
\noindent {\bf Graph-based Codecs and Attacks.} Recently, several software watermarking systems (or, equivalently, models) have been appeared in the literature that take or encode watermarks as graph structures and embed them into application programs. In general, such a graph-based software watermarking model mainly consists of two codec algorithms: an encoding algorithm which embeds a graph $G$ which represents a watermark $w$ into an application program $P$ resulting thus the watermarked program $P_w$, i.e., ${\tt embed}(P,G,k) \rightarrow P_w$, and a decoding algorithm which extracts the graph $G$ from $P_w$, i.e., ${\tt extract}(P_w, k) \rightarrow G$; we usually call the pair $({\tt embed}, {\tt extract})_G$ as {\it graph codec model} and the embedding and extracting algorithms as {\it codec} or {\it watermarking algorithms}.


Having designed a software watermarking algorithm, it is very important to evaluate it under various assessment criteria in order to gain information about its practical behavior \cite{CCKT03}; the most valuable and broadly used criteria can be divided into two main categories: (i) performance criteria (e.g., data-rate, time and space overhead, part protection, stealth, credibility), and (ii) resilience criteria (e.g., resistance against obfuscation, optimization, language-transformation) \cite{CT2002,MC06}. We mention that the performance criteria measure the behavior of the watermarked program $P_w$ and the quality and effectiveness of the embedded watermark $w$, while the resilience criteria measure the robustness and resistance of the embedded watermark $w$ against malicious user attacks.

From a graph-theoretical and practical point of view, we are interested in finding a class of graphs $\mathcal{G}$ having appropriate graph properties, e.g., graphs $G \in \mathcal{G}$ should contain nodes with small outdegree so that matching real program graphs, and developing software watermarking models $({\tt embed}, {\tt extract})_G$ which meet both:

\vspace*{-0.08in}
\begin{itemize}
\item[$\bullet$\,] {high performance:} both  programs, the original $P$ and the watermarked $P_w$, have almost identical execution behavior, almost same size and similar codes; and

\vspace*{-0.05in}
\item[$\bullet$\,] {high resiliency:} the algorithm ${\tt extract}()$ is insensitive to small changes of $P_w$ caused by various attacks, that is, if $G \in \mathcal{G}$ represents the watermark $w$ and ${\tt extract}(P_w,k) \rightarrow w$ then ${\tt extract}(P_w',k) \rightarrow w$ with $P_w' \approx P_w$.
\end{itemize}

\vspace*{-0.01in}
\noindent {\bf Related Work.} The most important software watermarking algorithms currently available in the literature are based on several techniques, among which the register allocation \cite{QP98}, spread-spectrum \cite{ZZZ11}, opaque predicate \cite{A02}, abstract interpretation \cite{CC04}, dynamic path techniques \cite{CCDHKLS2004}, code re-orderings \cite{Sharma11}; see also \cite{Book-CN10} for an exposition of the main results. It is worth noting that many algorithmic techniques on software watermarking have also received patent protection. \cite{Collberg11,DM96,Tarjan12,Rodriguez10}.

The patent of Davidson and Myhrvold \cite{DM96} presents the first published static software watermarking algorithm, which embeds the watermark into a program by reordering the basic blocks of a control flow-graph; note that a static watermark is stored inside programs' code in a certain format and it does not change during the programs' execution. Based on this idea, Venkatesan, Vazirani and Sinha \cite{VVS01} proposed the first graph-based software watermarking algorithm which embeds the watermark by extending a method's control flow-graph through the insertion of a directed subgraph; it is also a static algorithm called {\textmd VVS} or {\textmd GTW}. In \cite{VVS01}, the construction of the directed watermark graph $G$ is not discussed. Collberg et al.~\cite{CHCTS09} proposed an implementation of {\textmd GTW}, which they call {\textmd GTW$_{\textmd sm}$}, and it is the first publicly available implementation of the algorithm {\textmd GTW}. In {\textmd GTW$_{\textmd sm}$} the watermark is encoded as a reducible permutation graph (or, for short, RPG) \cite{CCKT03}, which is a reducible control flow-graph with a maximum out-degree of two, mimicking real code. Note that, for encoding integers the {\textmd GTW$_{\textmd sm}$} method uses only those permutations that are self-inverting. The first dynamic watermarking algorithm {\textmd CT} was proposed by Collberg and Thomborson \cite{CT99}; it embeds the watermark through a graph structure which is built on a heap at runtime.

Several software watermarking algorithms have been appeared in the literature that encode watermarks as graph structures \cite{CCKT03,CHCTS09,DM96,VVS01}. Recently, Chroni and Nikolopoulos extended the class of software watermarking codec algorithms and graph structures by proposing efficient and easily implemented algorithms for encoding numbers as reducible permutation flow-graphs (RPG) through the use of self-inverting permutations (or, for short, SiP). More precisely, they have presented an efficient method for encoding first an integer $w$ as a self-inverting permutation $\pi^*$ and then encoding $\pi^*$ as a reducible permutation flow-graph $F[\pi^*]$ \cite{CN10}; see also \cite{CN12}. The watermark graph $F[\pi^*]$ incorporates properties capable to mimic real code, that is, it does not differ from the graph data structures built by real programs. Moreover, the structural properties of $F[\pi^*]$ cause it resilient to edge, node and label modification attacks. Note that, the main idea of our dynamic watermarking model proposed in this work initially presented by Chroni and Nikolopoulos in \cite{CN12c}; see also \cite{CN13a,CN13b}.

\vspace*{0.1in}
\noindent {\bf Our Contribution.} In this paper, we present a dynamic watermarking model, which we call WaterRPG, for embedding the watermark graph $F[\pi^*]$ into an application program $P$ resulting thus the  watermarked program $P^*$. The main idea behind the proposed watermarking model is a systematic modification of appropriate function calls of the program $P$, through the use of control statements and opaque predicates, so that the execution of the watermarked program $P^*$ with a specific input gives a dynamic call-graph from which the watermark graph $F[\pi^*]$ can be easily constructed.
More precisely, for a specific input $I_{key}$ of a given program $P$, our model takes the dynamic call-graph $G(P, I_{key})$ of $P$ and the watermark graph $F[\pi^*]$, and produces the watermarked program $P^*$ so that the following key property holds: the dynamic call-graph $G(P^*, I_{key})$ of $P^*$ with input $I_{key}$ is isomorphic to the watermark graph $F[\pi^*]$. Within this idea the program $P^*$ is produced by only altering appropriate calls of specific functions of the input program $P$ and manipulating the execution flow of $P^*$ by including these altered function calls into control statements using opaque predicates. In the resulting watermarked program $P^*$, the control statements are executed following specific and well-defined execution rules and offer high functionality of $P^*$. Indeed, our model achieves low time and space overhead and ensures correctness, that is, $T(P,I)\approx T(P^*,I)$, $S(P,I)\approx S(P^*,I)$, and $O(P,I)= O(P^*,I)$ for every input $I$, where $T()$, $S()$, and $O()$ are the execution time, the heap space, and the output of $P$ or $P^*$ with input $I$. 

We have implemented our watermarking model WaterRPG on Java application programs downloaded from a free non
commercial game database, and evaluated its performance under various and commonly used watermarking evaluation criteria.
In particular, we selected a number of Java application programs and watermarked them using two main approaches: (i) the straightforward or
naive approach, and (ii) the stealthy approach. The naive approach watermarks a given program $P$ using the well-defined
call patterns of our model, while the stealthy approach watermarks $P$ using structural and programming properties of the code.

The evaluation results show the efficient functionality of all the Java programs $P^*$ watermarked under both the naive and
stealthy cases. The experiments also show that the watermarking approaches supported by our model can help develop efficient
watermarked Java programs with respect to time and space overhead, credibility, stealthiness, and other watermarking metrics. Moreover, our WaterPRG model incorporates properties which cause it resilient to several watermark and code attacks.

Table~\ref{tab:properties} summarizes the most important general properties, in complementary or opposite pairs, of a software watermarking model and shows the properties of our WarerRPG model.
Throughout the paper, for a given program $P$ we shall denote by $P^*$ the watermarked program produced by our model WaterRPG.

\begin{table}[t!]
\begin{center}
\begin{tabular}{| c | c |}
\hline
{Models' Properties}                    &{WaterRPG's Properties}                        \\
\hline \hline
{static - dynamic}                      &{\ dynamic (execution trace) \ }                         \\
{robust - fragile}                      &{robust}                         \\
{visible - invisible}                   &{invisible}                         \\
{blind - informed}                      &{blind}                         \\
{\ focus - spread spectrum \ }          &{spread spectrum}                         \\
\hline
\end{tabular}
\vspace{0.06cm}
\caption{General properties of watermarking models and the properties of our WarerRPG model.}
\label{tab:properties}
\end{center}
\end{table}

\vspace*{0.1in}
\noindent {\bf Road Map.} The paper is organized as follows: In Section~2 we establish the notation and related terminology, and present background results. In Section~3 we present our dynamic watermarking model WaterRPG; we first describe its structural and operational components and then the embedding algorithm ${\tt  Encode\_RPG.to.CODE}$ and the extracting algorithm ${\tt Decode\_CODE.to.RPG}$. In Section~4 we implement our watermarking model in real Java application programs and show two main watermarking approaches supported by the WaterRpg model, namely naive and stealthy. In Section~5 we evaluate our model under several software watermarking assessment criteria, while in Section~6 we summarize our work and propose possible future extensions.

\section{Background Results}
\label{sec:Background Results}

\noindent In this section, we present background results and key objects that are used in the design of our watermarking model WaterRPG. In particular, we briefly present the main results of our previous work concerning the process of encoding numbers as graph structures namely reducible permutation graphs (or, for short, RPG); we denote such a graph as $F[\pi^*]$. We also briefly discuss properties of dynamic call-graphs which are used as key-objects in our watermarking model for embedding the graph $F[\pi^*]$ into an application program.

\vspace*{0.1in}
\subsection{Encode Numbers as RPGs}
\label{Encode-Numbers}

\noindent We consider finite graphs with no multiple edges. For a graph~$G$, we denote by $V(G)$ and $E(G)$ the vertex set and edge set of $G$, respectively. We also consider permutations over the set $N_n = \{1, 2, \ldots, n\}$.

Let $\pi$ be a permutation over the set $N_n$. We think of permutation $\pi$ as a sequence $(\pi_1, \pi_2, \ldots, \pi_n)$, so, for example, the permutation $\pi = (4,7,6,1,5,3,2)$ has $\pi_1 = 4$, $\pi_2 = 7$, etc. Notice that $\pi^{-1}_i$ is the position in the sequence of the number $i$; in our example, $\pi_4^{-1} = 1$, $\pi_1^{-1} = 4$, $\pi_3^{-1} = 6$, etc.
The inverse of a permutation $\pi$ is the permutation $\tau=(\tau_1, \tau_2, \ldots, \tau_n)$ with $\tau_{\pi_i} = \pi_{\tau_i} = i$. A {\it self-inverting permutation} (or, involution) is a permutation that is its own inverse, i.e., $\pi_{\pi_i} = i$. Throughout the paper we denote a self-inverting permutation $\pi$ (or, for short, SiP) over the set $N_n$ as $\pi^*$.

A flow-graph is a directed graph $F$ with an initial node $s$ from which all other nodes are reachable. A directed graph $G$ is strongly connected when there is a path $x \rightarrow y$ for all nodes $x$, $y$ in $V(G)$. A node $x$ is an {\it entry} for a subgraph $H$ of the graph $G$ when there is a path $p = (y_1, y_2, \ldots, y_k, x)$ such that $p \cap H = \{x\}$.
A flow-graph is reducible when it does not have a strongly connected subgraph with two (or more) entries. There are at least three other equivalent definitions; see, \cite{HU74}.

\vspace*{0.2in}
\noindent {\bf (I) Encode Numbers as SiPs}

\vspace*{0.1in}
\noindent In \cite{CN10}, we introduced the notion of {\it bitonic permutations} and we presented two algorithms, namely ${\tt Encode\_W.to.SiP}$ and ${\tt Decode\_SiP.to.W}$, for encoding an integer $w$ into a self-inverting permutation $\pi^*$ and extracting it from $\pi^*$. We have actually proved the following results.

\vspace*{0.1in}
\noindent {\bf Theorem~2.2.}
{\it Let $w$ be an integer and let $b_1b_2\cdots b_n$ be the binary representation of $w$. The algorithm {\textmd Encode\_W.to.SiP} encodes the number $w$ in a self-inverting permutation $\pi^*$ of length $2n+1$ in $O(n)$ time and space.}

\vspace*{0.1in}
\noindent {\bf Theorem~2.3.}
{\it Let $\pi^*$ be a self-inverting permutation of length $n$ which encodes an integer $w$ using the algorithm {\textmd Encode\_W.to.SiP}. The algorithm {\textmd Decode\_SiP.to.W} correctly decodes the permutation $\pi^*$ in $O(n)$ time and space.}

\vspace*{0.2in}
\noindent {\bf (II) Encode SiPs as RPGs}

\vspace*{0.1in}
\noindent Recently, we have presented an efficient and easily implemented algorithm for encoding numbers as reducible permutation flow-graphs through the use of self-inverting permutations \cite{CN12}. In particular, we have proposed the algorithm ${\tt Encode\_SiP.to.RPG}$, which encodes a SiP $\pi^*$ as a reducible permutation flow-graph $F[\pi^*]$ by exploiting domination relations on the elements of $\pi^*$ and using an efficient DAG representation of $\pi^*$. 

\begin{figure}[t]
\hrule \medskip
  \medskip
  \centering
  \includegraphics[scale=0.50]{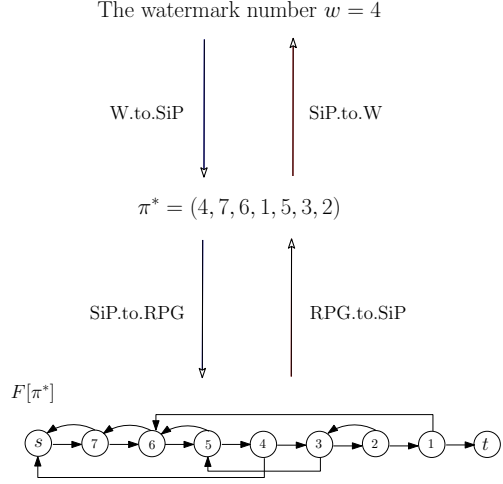}
  \centering
  \medskip \hrule\medskip\medskip
\caption{\small{The main data components used by our codec algorithms (i.e., the watermark $w$, the SiP $\pi^*$, and the RPG $F[\pi^*]$) and a flow of the process of encoding a watermark number $w$ into the graph $F[\pi^*]$ and extracting it from $F[\pi^*]$.}}
\label{components}
\end{figure}

The whole encoding process takes $O(n)$ time and requires $O(n)$ space, where $n$ is the length of the self-inverting permutation $\pi^*$.
The decoding process takes time and space linear in the size of the flow-graph $F[\pi^*]$, that is, the algorithm ${\tt Decode\_RPG.to.SiP}$ takes $O(n)$ time and space. Our results presented in \cite{CN12} are summarized in the following theorems.

\newpage

\vspace*{0.1in}
\noindent {\bf Theorem~2.4.}
{\it Let $\pi^*$ be a self-inverting permutation over the set $N_n$. The algorithm {\textmd Encode\_SiP.to.RPG} encodes the permutation $\pi^*$ into a reducible permutation graph $F[\pi^*]$ in $O(n)$ time and space.}

\vspace*{0.1in}
\noindent {\bf Theorem~2.5.}
{\it Let $F[\pi^*]$ be a reducible permutation graph of order $O(n)$ produced by the encoding algorithm {\textmd Encode\_SiP.to.RPG}. The algorithm {\textmd Encode\_RPG.to.SiP} correctly extracts the permutation $\pi^*$ from $F[\pi^*]$ in $O(n)$ time and space.}

\vspace*{0.1in}
\noindent The reducible permutation graph $F[\pi^*]$ of the self-inverting permutation $\pi^*$ is directed with a descending ordering on its nodes $s=u_{n+1}$, $u_{n}$, $\dots$, $u_1, u_0=t$. Hereafter, we shall call the edge $(u_i, u_j)$ of graph $F[\pi^*]$ {\it forward} if $i > j$ and {\it backward} otherwise.

Figure~\ref{components} depicts the main data components used by our codec algorithms, i.e., the watermark number $w$, the self-inverting permutation $\pi^*$, and the reducible permutation flow-graph $F[\pi^*]$. The same figure shows a flow of the process of encoding a watermark number $w$ into the graph $F[\pi^*]$ and extracting it from $F[\pi^*]$.

\vspace*{0.1in}
\subsection{Call-graphs}
\noindent A {\it call-graph} is a directed graph that represents calling relationships between program units in a computer program. Specifically, the nodes $f_1, f_2, \ldots, f_n$ of a call-graph represent functions, procedures, classes, or similar program units and each edge $(f_i, f_j)$ indicates that $f_i$ calls $f_j$; function $f_i$ is called {\it caller} while $f_j$ is called {\it callee}.

Call-graphs can be divided in two main classes of graphs, namely {\it static} and {\it dynamic}. A static call-graph is the structure describing those invocations that could be made from one program unit to another in any possible execution of the program \cite{XN2002}. The static call-graph can be determined from the program source code; we mention that, its construction is a time consuming process specifically in the case of large scale software \cite{GKM1982}.

A dynamic call-graph $G$ is a directed graph that includes invocations of caller--callee pairs over an execution of the program $P$.
Such a graph can be considered as an instance of the corresponding static call-graph for a specific input sequence $I$.
The call-graph $G$ is a data structure that is used by dynamic optimizers for analyzing and optimizing the whole-program's behavior; such a graph can be extracted by a profiler. It is fair to mention that the construction of a dynamic call-graph $G$ of a program $P$ is not a time consuming process even if $P$ is a large scale software.

Throughout the paper we denote a dynamic call-graph $G$ of the program $P$ over the input $I$ as $G(P,I)$. Figure~\ref{call-graphs}(a) depicts the structure of the dynamic call-graph $G(P,I_{key})$ of an application program $P$ with input $I_{key}$.

\vskip 0.3in 
\section{The Dynamic Watermarking Model}
\label{sec:Watermarking Model}
\noindent Having encoded a watermark number $w$ as reducible permutation graph $F[\pi^*]$, let us now present our dynamic watermarking model WaterRPG; we first demonstrate its structural and operational components and, then, we describe the embedding and extracting watermarking algorithms.

\vspace*{0.1in}
\subsection{Operational Framework}
The main idea behind the proposed watermarking model is a systematic modification of appropriate function calls of the program $P$ so that the execution of the resulting watermarked program $P^*$ with a specific input $I_{key}$ gives a dynamic call-graph $G(P^*, I_{key})$ from which the watermark graph $F[\pi^*]$ can be easily constructed.

More precisely, the main operations performed by the WaterRPG model can be outlined as follows: for a specific input $I_{key}$ of the original program $P$, it takes the dynamic call-graph $G(P, I_{key})$ and the graph $F[\pi^*]$, and produces the watermarked program $P^*$ so that its dynamic call-graph $G(P^*, I_{key})$ with input $I_{key}$ is isomorphic to the watermark graph $F[\pi^*]$. The call-graphs $G(P, I_{key})$ and $G(P^*, I_{key})$ dictate the execution flow of the original program $P$ and the watermarked program $P^*$, respectively. Thus, since the call-graph $G(P, I_{key})$ is not isomorphic to $G(P^*, I_{key})$ in general, the model controls the flow of selected function calls of $P^*$ so that $O(P,I)=O(P^*,I)$ for every input $I$, where $O(P,I)$ (resp. $O(P^*,I)$) is the output of the program $P$ (resp. $P^*$) with input $I$. In this framework, the program $P^*$ is produced by only altering appropriate calls of specific functions of the input program $P$.

\begin{figure}[t]
  \centering
  \hrule
  \medskip\medskip
  \includegraphics[scale=0.60]{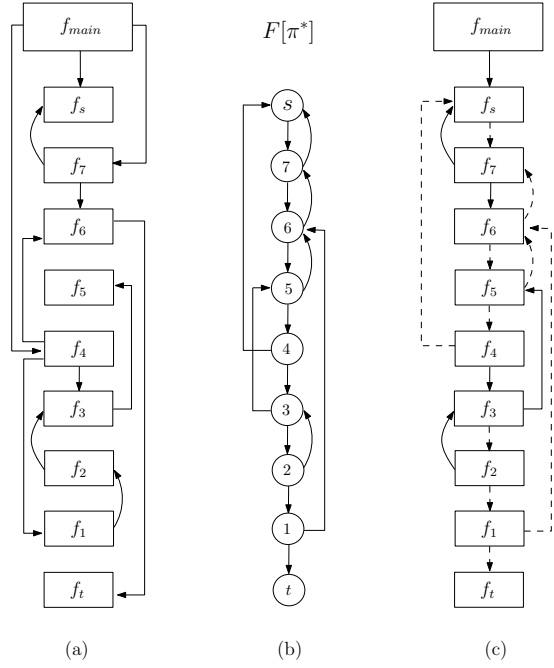}
  \medskip
  \hrule
  \centering\medskip\medskip
  \caption{\small{(a) The dynamic call-graph $G(P, I_{key})$ of an application program $P$. (b) The reducible permutation graph $F[\pi^*]$. (c) The dynamic call-graph $G(P^*, I_{key})$ of the watermarked program $P^*$.}}
  \label{call-graphs}
\end{figure}

Figure~\ref{call-graphs} shows the dynamic call-graph $G(P, I_{key})$ of an application program $P$, the reducible permutation graph $F[\pi^*]$ which encodes the number $w=4$ and the dynamic call-graph $G(P^*, I_{key})$ of the watermarked program $P^*$.

\vspace*{0.1in}
\subsection{Model Components}
\label{Model Components}
\noindent Our watermarking model uses two main categories of components namely {\it structural} components and {\it operational} components. The first category includes the dynamic call-graph $G(P, I_{key})$ of the input program $P$, the watermark graph $F[\pi^*]$, and the dynamic call-graph $G(P^*, I_{key})$ of the watermarked program $P^*$, while the second category includes call patterns, control statements and execution rules which are components related to the process of embedding the graph $F[\pi^*]$ into application program $P$.

We next describe the construction and main properties of the dynamic call-graph $G(P^*, I_{key})$, two call patterns based on which we correspond edges of the call-graph $G(P^*, I_{key})$ to function calls, and specific variables and statements which control the execution of real and water functions.

\vspace*{0.2in}
\noindent {\bf (I) The Dynamic Call-graph G(P$^*$, I$_{key}$)}

\vspace*{0.1in}
\noindent Let $F[\pi^*]$ be a watermark graph (or, equivalently, water-graph) on $n+2$ nodes and $G(P, I_{key})$ be the dynamic call-graph of a program $P$ on $n+3$ nodes $f_{main}, f_s, f_1, \ldots, f_{n}, f_{t}$ taken after running the program $P$ with the input $I_{key}$. In general, the selection of the input $I_{key}$ is such that it produces the call-graph $G(P, I_{key})$ having structure as ``close" as possible to the structure of $F[\pi^*]$. We assign the $n+2$ nodes $f_s=f_{n+1}, f_n, \ldots, f_{1}, f_0=f_t$ of the call-graph $G(P, I_{key})$ to $n+2$ nodes $s=u_{n+1}, u_n, \ldots, u_1, u_{0}=t$ of $F[\pi^*]$ into 1-1 correspondence; the main function $f_{main}$ do not correspond to any node of $F[\pi^*]$.

Let $(u_i, u_j)$ be an edge in graph $F[\pi^*]$ and let $(f_i, f_j)$ be an edge in call-graph $G(P, I_{key})$. We say that the edge $(f_i, f_j)$ corresponds to edge $(u_i, u_j)$ iff the node $f_i$ corresponds to $u_i$ and the node $f_j$ corresponds to $u_j$, $0 \leq i, j \leq n+1$. Moreover, if $(u_i, u_j)$ is a forward (resp. backward) edge in the graph $F[\pi^*]$ we say that the corresponding edge $(f_i, f_j)$ in graph $G(P, I_{key})$ is a forward (resp. backward) edge.

The dynamic call-graph $G(P^*, I_{key})$ is constructed as follows:
\begin{itemize}
  \item[$\bullet$] $V(G(P^*, I_{key}))=V(G(P, I_{key}))$, i.e., it has the same nodes as the call-graph $G(P, I_{key})$;
  \item[$\bullet$] $E(G(P^*, I_{key}))=E(F[\pi^*])$, i.e., $(f_i, f_j)$ is an edge in $E(G(P^*, I_{key}))$ iff the corresponding $(u_i, u_j)$ is an edge in $F[\pi^*]$.
\end{itemize}

\noindent The edges of the call-graph $G(P^*, I_{key})$ are divided into two categories namely {\it real} and {\it water} edges; note that, the real (resp. water) edges correspond to real (resp. water) function calls. An edge $(f_i, f_j)$ of the call-graph $G(P, I_{key})$ is characterized as either

\begin{itemize}
  \item[$\bullet$] {\it real edge} if $(f_i, f_j)$ is an edge in $G(P, I_{key})$, or
  \item[$\bullet$] {\it water edge} if $(f_i, f_j)$ is not an edge in $G(P, I_{key})$.
\end{itemize}

\noindent Figure~\ref{call-graphs}(c) shows the dynamic call-graph $G(P^*, I_{key})$ along with its real edges (solid arrows) and water edges (dashed arrows).

\vspace*{0.2in}
\noindent {\bf (II) Call Patterns}

\vspace*{0.1in}
\noindent In the implementation phase, we modify the source code of program $P$ using specific function call patterns which we describe below.

Let $P$ be an application program, $G(P, I_{key})$ be the dynamic call-graph of the program $P$ with input $I_{key}$, and $F[\pi^*]$ be a watermark-graph which we have to embed into $P$. According to our watermarking model, the embedding process relies mainly on altering the execution-flow of appropriate function calls of $P$ such that the execution of the resulting program $P^*$ with the input $I_{key}$ produces a
call-graph $G(P^*, I_{key})$ which, after removing the node $f_{main}$, is isomorphic to watermark-graph $F[\pi^*]$.

Let $(f_i, f_j)$ be an edge of call-graph $G(P^*, I_{key})$ or, equivalently, an edge which we want to appear in $G(P^*, I_{key})$. Since $G(P^*, I_{key})$ has two types of edges it follows that $(f_i, f_j)$ is either real or water edge. Based on the type of $(f_i, f_j)$, we do the following:
\begin{itemize}
  \item[$\bullet$] if $(f_i, f_j)$ is a water edge we add the statement ${\tt call}(f_j)$ in the function $f_i$, while
  \item[$\bullet$] if $(f_i, f_j)$ is a real edge we add no call statement since the statement ${\tt call}(f_j)$ exists in $f_i$.
\end{itemize}

\begin{figure}[!t]
    \centering
    \hrule
        \medskip\medskip
        \subfigure[]{\includegraphics[scale=0.55]{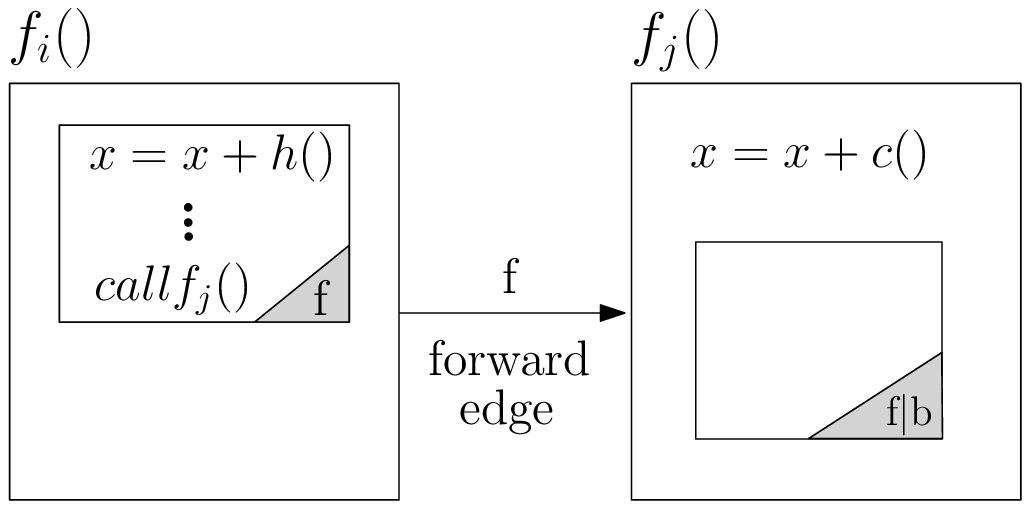}}
        \quad\quad\quad\quad\phantom{.}
        \subfigure[]{\includegraphics[scale=0.55]{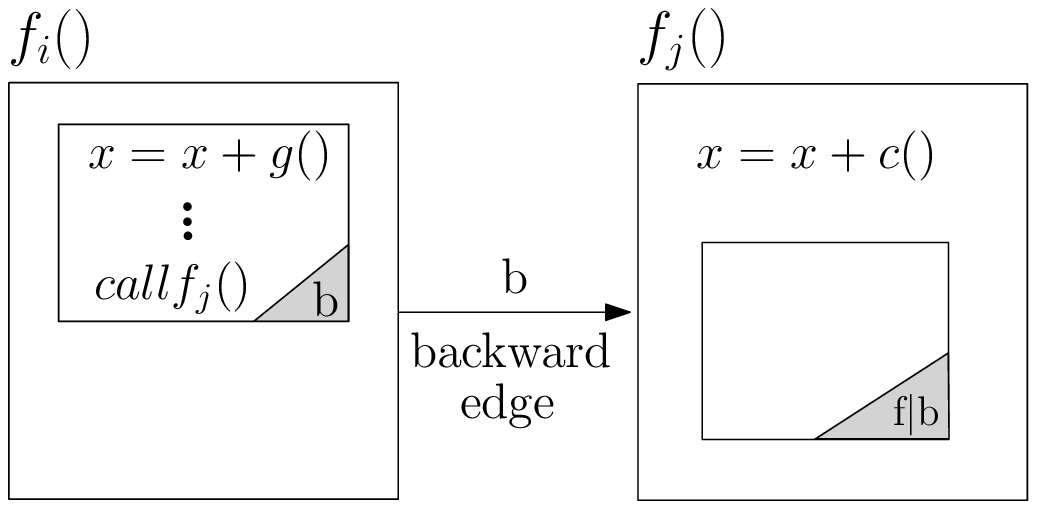}}
        \quad\quad\quad
        \subfigure[]{\includegraphics[scale=0.74]{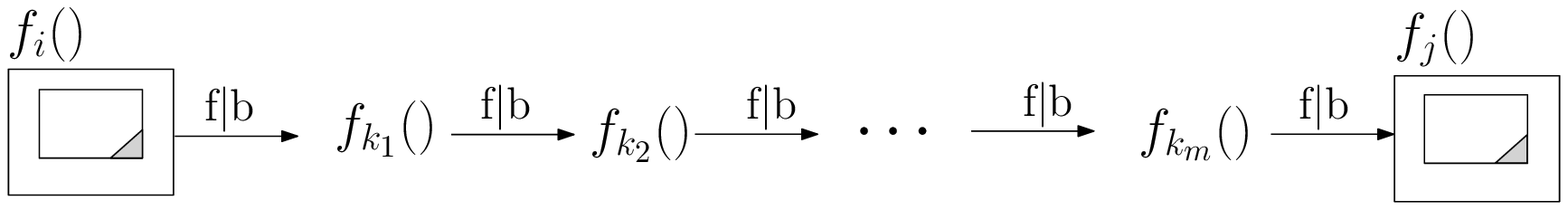}}
    \medskip
    \hrule
    \centering\medskip\medskip
    \caption{\small{(a) The forward call pattern {\it f-call}; (b) The backward call pattern {\it b-call};
    (c) The path call pattern {\it p-call}. The boxes with marked corners are the f$|$b-blocks.}}
    \label{call-patterns}
\end{figure}

\noindent Based on whether $(f_i, f_j)$ is either a forward or a backward edge we add specific statements in functions $f_i$ and $f_j$ according to the following two call patterns namely {\it forward} and {\it backward} call patterns:

\begin{itemize}
  \item[(a)] if $(f_i, f_j)$ is a forward edge we add the statement $x=x+h()$ in function $f_i$ before the call-site or, equivalently, call-point of the function $f_j$, and the statement $x=x+c()$ in the function $f_j$, while
  \item[(b)] if $(f_i, f_j)$ is a backward edge we add the statement $x=x+g()$ in function $f_i$ before the call-site of the function $f_j$, and the statement $x=x+c()$ in the function $f_j$,
\end{itemize}

\noindent where $x$ is a variable and $h()$, $g()$ and $c()$ are functions. Figure~\ref{call-patterns}(a) depicts the forward call pattern or, for short, {\it f-call}, while Figure~\ref{call-patterns}(b) depicts the backward call pattern or, for short, {\it b-call}.

Recall that the direct edge $(f_i, f_j)$ of a call-graph represents a function call operation where $f_i$ is the caller function and $f_j$ the callee function; in other words, it means that in function $f_i$ there exists the statement ${\tt call}$($f_j$). Hereafter, in this case we shall say that $(f_i, f_j)$ is a direct call.

In a call-graph of an application program we usually meet sequences of calls of the form $(f_i, f_{k_1}, f_{k_2}, \ldots, f_{k_m}, f_j)$. For simplicity we set $f_i=f_{k_0}$ and $f_j=f_{k_{m+1}}$ and suppose that each of these calls $(f_{k_0}, f_{k_1})$, $(f_{k_1}, f_{k_2})$, $\ldots$, $(f_{k_m}, f_{m+1})$ is either forward or backward. We extend the notion of the direct call $(f_i, f_j)$ to indirect call $(f_i \rightarrow f_j)$; an indirect call consists of a path of functions $(f_i, f_{k_1}, \ldots, f_j)$ of length $\ell \geq 2$. Using the f-call and b-call patterns, we next define the path call pattern or, for short, {\it p-call} as follows:

\begin{itemize}
  \item[(c)] if $(f_{k_{i}}, f_{k_{i+1}})$ and $(f_{k_{i+1}}, f_{k_{i+2}})$ are two consecutive calls of a call sequence, we apply an f-call or a b-call in $(f_{k_{i+1}}, f_{k_{i+2}})$ by first adding either the statement $x=x+h()$ or $x=x+g()$ in function $f_{k_{i+1}}$ after the call-point of statement $x=x+c()$, and then adding the statement $x=x+c()$ in $f_{k_{i+2}}$, $0 \leq i \leq m-1$.
\end{itemize}

\noindent Figure~\ref{call-patterns} shows the structures of the patterns f-call and b-call of the direct call $(f_i, f_j)$, and the structure of the pattern p-call of an indirect call $(f_i \rightarrow f_j)$.

\begin{figure}[t]
  \centering
  \hrule
  \medskip\medskip
  \includegraphics[scale=0.50]{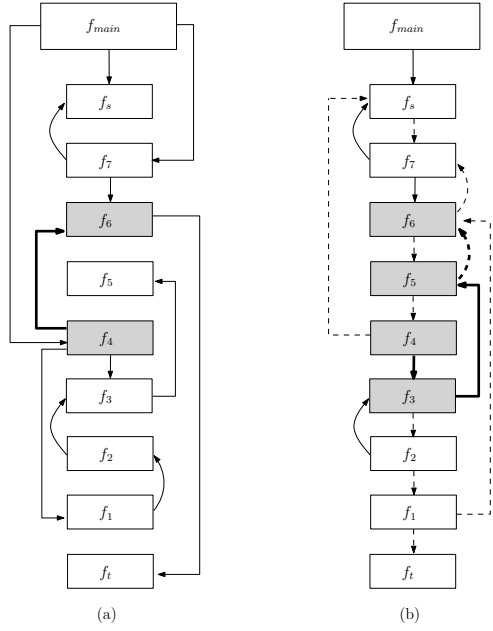}
  \medskip
  \hrule
  \centering\medskip\medskip
  \caption{\small{(a) The real-call $(f_4,f_6)$ in the call-graph $G(P, I_{key})$ of a program $P$; bold arrow. (b) The corresponding path-call $(f_4, f_3, f_5, f_6)$ in the call-graph $G(P^*, I_{key})$ of the watermarked program $P^*$; bold arrows.}}
\label{path}
\end{figure}

\vspace*{0.2in}
\noindent {\bf (III) Control Statements and Variables}

\vspace*{0.1in}
\noindent In any watermarking model both the original program $P$ and the watermarked program $P^*$ have to operate identically. Thus, since the call-graphs $G(P, I_{key})$ and $G(P^*, I_{key})$ dictate the execution flow of programs $P$ and $P^*$, respectively, and the call-graph $G(P, I_{key})$ is not isomorphic to $G(P^*, I_{key})$, we have to control the flow of selected function calls of program $P^*$ so that $O(P,I)=O(P^*,I)$ for every input $I$.

To do this, we exploit the values of specific variables in a function $f_i$ by using them in some selected or added control statements as part of opaque predicates. More precisely, we use the variable $x$ of the f-call and b-call patterns and include it in a specific control statement $s$ causing thus an ``appropriate execution flow" of the functions of the call-graph $G(P^*, I_{key})$; with the term ``appropriate execution flow" we mean that the execution flow of the functions of the call-graph $G(P^*, I_{key})$ is such that $O(P,I)=O(P^*,I)$ for every input $I$. Hereafter, we call {\it cf-statement} the control statement $s$ and {\it cf-variable} the variable $x$. In this point, we also define the {\it f-block} and {\it b-block} to be specific parts of the code which contain (i) cf-statements, (ii) cf-variables, and (iii) water-foreword or water-backward function calls. We denote by f$|$b-block either an f-block or a b-block; in Figure~\ref{call-patterns}, the f$|$b-blocks are shown by boxes with marked corners.

We next describe the mechanism which ensures an appropriate execution flow of the functions of $G(P^*, I_{key})$ through the altering of the execution flow of the functions of the program $P$ by modifying or adding some specific control statements. In fact, what the mechanism actually does is to modify the conditions or expressions of these control statements by adding opaque predicates.

\vspace*{0.1in}
\noindent {\bf Definition~3.1.} A predicate $Q$ is opaque at a program point $p$, if at point $p$ the outcome of $Q$ is known at embedding time. If $Q$ always evaluates to \emph{true} we write $Q_p^{T}$, for \emph{false} we write $Q_p^{F}$, and if $Q$ sometimes evaluates to \emph{true} and sometimes to \emph{false} we write $Q_p^{?}$.

\vspace*{0.1in}
\noindent Let $(f_i, f_j)$ be a direct call in our program $P^*$ or, equivalently, an edge in the call-graph $G(P^*, I_{key})$; it is either real-forward, real-backward, water-forward, or water-backward edge. In any case, the proposed mechanism uses the value of the cf-variable $x$ and makes the following operations:

\begin{itemize}
  \item[$\bullet$] in function $f_i$: create a control statement (if, switch, for, while, etc), add an opaque predicate $Q_p^{?}$ containing the cf-variable $x$, and insert it at a point $p$ before the statement $x=x+h()$ or $x=x+g()$; we could also select a control statement at a point $p$, if there exists, consider it as cf-statement and include the opaque predicate $Q_p^{?}$ in its condition part.

  \item[$\bullet$] in function $f_j$: create a control statement as in function $f_i$, if such a statement does not exist, and insert it at a point $p$ before the statement $x=x+c()$; if such a statement exists, we only add a new opaque predicate $Q_p^{?}$ in the condition of that statement. The main body of $f_j$ is included in a block of a cf-statement the execution of which is depending upon the behavior (i.e., true or false) of the opaque predicate $Q_p^{?}$.
\end{itemize}

\newcommand*{\MyInd}{\hspace*{0.3cm}}
\newcommand*{\MyIndent}{\hspace*{0.6cm}}
\begin{figure}[t!]

\begin{center}
\begin{tabular}{|p{3.5cm} | p{3.5cm}|}
\hline
{Program $P$}                   &{Program $P^*$}                        \\
\hline \hline
{${\tt function}$ $f_i()$}                  &{${\tt function}$ $f_i()$}                         \\
{\MyInd \ldots}                     &{\MyInd \ldots}                            \\
{}                                  &{\MyInd ${\tt if}$ (\it{condition} \& $Q_p^{?}$)}     \\
{\MyInd ${\tt if}$ (\it{condition})}        &{\MyIndent \ldots}                         \\
{\MyIndent \ldots}                  &{\MyIndent ${\tt x = x + h()}$;}                 \\

{\MyIndent ${\tt statements}$;}             &{\MyIndent \ldots}                         \\
{\MyIndent \ldots}                  &{\MyIndent  \textbf{call $f_j()$};}        \\
{}                                  &{\MyIndent  \ldots}                        \\
{}                                  &{\MyIndent  ${\tt statements}$;}                   \\
{}                                  &{\MyIndent \ldots}                         \\
\hline
\end{tabular}
\vspace{0.1cm}
\caption{An example of cf-statement modification via opaque predicates in the case where $(f_i, f_j)$ is a water-forward function call.}
\label{tab:water-forward}
\end{center}
%
\vspace{0.2cm}
%
\begin{center}
\begin{tabular}{|p{3.5cm} | p{3.5cm}|}
\hline
{Program $P$}                   &{Program $P^*$}                            \\
\hline \hline
{${\tt function}$ $f_i()$}                  &{${\tt function}$ $f_i()$}                             \\
{\MyInd \ldots}                     &{\MyInd \ldots}                                \\
{}                                  &{\MyInd ${\tt if}$ (\it {condition} \& $Q_p^{?}$)}  \\
{\MyInd ${\tt if}$ (\it {condition})}      &{\MyIndent \ldots}                             \\
{\MyIndent \ldots}                  &{\MyIndent ${\tt x = x + g()}$;}                     \\

{\MyIndent\textbf{call $f_j()$};}  &{\MyIndent \ldots}                             \\
{\MyIndent \ldots}                  &{\MyIndent \textbf{call $f_j()$};}            \\
{\MyIndent  ${\tt statements}$;}            &{\MyIndent  \ldots}                            \\
{\MyIndent \ldots}                  &{\MyIndent  ${\tt statements}$;}                       \\
{}                                  &{\MyIndent  \ldots}                            \\
\hline
\end{tabular}
\vspace{0.1cm}
\caption{An example of cf-statement modification via opaque predicates in the case where $(f_i, f_j)$ is a real-backward function call.}
\label{tab:real-backward}
\end{center}
%
\vspace{0.2cm}
%
\begin{center}
\begin{tabular}{|p{3.5cm} | p{3.5cm}|}
\hline
\multicolumn{2}{|c|}{Call $(f_i, f_j)$ of Program $P^*$}                            \\
\hline \hline
{${\tt function}$ $f_i()$}                                    &{${\tt function}$ $f_j()$}                             \\
{\MyInd \ldots}                                       &{\MyInd \ldots}                                \\
{\MyInd ${\tt if}$ (\it {condition} \& $Q_p^{?}$)}          &{\MyInd ${\tt if}$ (\it {condition} \& $Q_p^{?}$)}  \\
{\MyIndent \ldots}                                    &{\MyIndent \ldots}                             \\
{\MyIndent ${\tt x = x + h()}$;}                              &{\MyIndent ${\tt x = x + c()}$;}                     \\
{\MyIndent \ldots}                                    &{\MyIndent \ldots}                             \\
{\MyIndent \textbf{call $f_j()$};}                    &{\MyInd ${\tt if}$ (\it {condition} \& $Q_p^{?}$)}            \\
{\MyIndent  \ldots}                                   &{\MyIndent  \ldots}                            \\
{\MyIndent  ${\tt statements}$;}                              &{\MyIndent  ${\tt statements}$;}                       \\
{\MyIndent  \ldots}                                   &{\MyIndent  \ldots}                            \\
\hline
\end{tabular}
\vspace{0.1cm}
\caption{An example of cf-statement modification via opaque predicates of the function $f_j$ in the case where $(f_i, f_j)$ is a water-forward function call.}
\label{tab:caller-callee}
\end{center}
\end{figure}

\noindent Note that, the above operations form specific parts of code of functions $f_i$ and $f_j$, namely f$|$b-blocks, i.e., either f-blocks or b-blocks; see, Figure~\ref{call-patterns}.

Figure~\ref{tab:water-forward} shows an example of the modification of the condition part of an ${\tt if}$ cf-statement via an opaque predicate; since $(f_i, f_j)$ is a water-forward function call, the statement ${\tt call}$$(f_j)$ does not exist in function $f_i$, and thus we add it in $f_i$, while the cf-statement is the $x = x + h()$. On the other hand, Figure~\ref{tab:real-backward} shows an example in case where $(f_i, f_j)$ is a real-backward function call. In this case, the statement ${\tt call}$$(f_j)$ does exist in $f_i$ while the cf-statement is the $x = x + g()$. Figure~\ref{tab:caller-callee} shows an example of the modification of the function $f_j$ in the case where $(f_i, f_j)$ is a water-forward function call.

\vspace*{0.12in}
\noindent {\bf Remark~3.1.} Based on the structural properties of the watermark graph $F[\pi^*]$ and call-graph $G(P^*, I_{key})$ we can easily prove the following lemma.

\vspace*{0.1in}
\noindent {\bf Lemma~3.1.}
{\it Let $G(P, I_{key})$ and $G(P^*, I_{key})$ be the call-graphs of programs $P$ and $P^*$, respectively, on input $I_{key}$, and let $(f_i, f_j)$ be an edge in call-graph $G(P, I_{key})$. Then, there always exists an edge $(f_i, f_j)$ or a path $(f_i, f_{k_1}, f_{k_2}, \ldots, f_{k_m}, f_j)$ in call-graph $G(P^*, I_{key})$.}

\vspace*{0.1in}
\noindent {\bf Remark~3.2.} In our implementation, in the case where $(f_i, f_j)$ is an edge in $G(P, I_{key})$ and $(f_i, f_j)$ is not an edge in $G(P^*, I_{key})$ we have to compute a path $(f_i, f_{k_1}, \ldots, f_j)$ of function calls in $G(P^*, I_{key})$. Such a path is a shortest path from $f_4$ to $f_6$ in the graph $G(P^*, I_{key})$; it may consist of all types of edges, that is, real-forward or real-backward and water-forward or water-backward edges. Figure~\ref{path}(a) shows the edge $(f_4,f_6)$ in $G(P, I_{key})$ which is not an edge in $G(P^*, I_{key})$, while Figure~\ref{path}(b) shows its corresponding shortest path from $f_i$ to $f_j$, that is, the path $(f_4,f_3,f_5,f_6)$; note that, $(f_4,f_3)$ is a real-forward edge, $(f_3,f_5)$ is a real-backward edge, and $(f_5,f_6)$ is a water-backward edge.

\vspace*{0.2in}
\noindent {\bf (VI) Execution Rules}

\vspace*{0.1in}
\noindent We present the rules based on which we control the execution flow of the functions of $P^*$ such that $O(P,I)=O(P^*,I)$ for every input $I$. In fact, we show in all the cases how the value of $Q_p^{?}$ dictates the execution flow of functions of $G(P^*, I_{key})$.

Let $(f_i, f_j)$ be a direct call in program $P^*$ or, equivalently, an edge in the call-graph $G(P^*, I_{key})$. We distinguish the following cases:

\vspace*{0.02in}
\begin{itemize}
  \item[$\bullet$] $(f_i, f_j)$ is {\bf real-forward} or {\bf real-backward}: in this case we modify the functions $f_i$ and $f_j$ as follows:
    \begin{itemize}
        \item[$\circ$] Function $f_i$: the opaque predicate $Q_p^{?}$ in the cf-statement before the cf-value $x=x+h()$ or $x=x+g()$ and the ${\tt call}$($f_j$) is evaluated to true, that is, $Q_p^{T}$.
        \item[$\circ$] Function $f_j$: the opaque predicate $Q_p^{?}$ in the cf-statement before the cf-value $x=x+c()$ is evaluated to true, that is, $Q_p^{T}$, while the $Q_p^{?}$ for the cf-statement which controls the statements of the main body of the function $f_j$ is also evaluated to true, that is, $Q_p^{T}$.
    \end{itemize}

    \vspace*{0.02in}
    \item[$\bullet$] $(f_i, f_j)$ is {\bf water-forward} or {\bf water-backward}: in this case we modify the functions $f_i$ and $f_j$ as follows:
    \begin{itemize}
        \item[$\circ$] Function $f_i$: the opaque predicate $Q_p^{?}$ in the cf-statement before the cf-value $x=x+h()$ or $x=x+g()$ and the ${\tt call}$$(f_j)$ is evaluated to true, that is, $Q_p^{T}$.
        \item[$\circ$] Function $f_j$: the opaque predicate $Q_p^{?}$ in the cf-statement before the cf-value $x=x+c()$ is evaluated to true, that is, $Q_p^{T}$, while the $Q_p^{?}$ for the cf-statement which controls the statements of the main body of the function $f_j$ is evaluated to false, that is, $Q_p^{F}$.
    \end{itemize}
\end{itemize}

\vspace*{0.1in}
\noindent Recall that a predicate $Q$ is opaque at a program point $p$, if at point $p$ the outcome of $Q$ is known at the embedding time. 

\vspace*{0.1in}
\noindent {\bf Remark~3.3.} During the execution of the function $f_i$ of the program $P^*$ only one opaque predicate $Q_p^{?}$ of the cf-statements is evaluated to true with respect to the current value of the cf-variable $x$.

\begin{figure}[!t]
    \centering
    \hrule
        \medskip\medskip
        \includegraphics[scale=0.65]{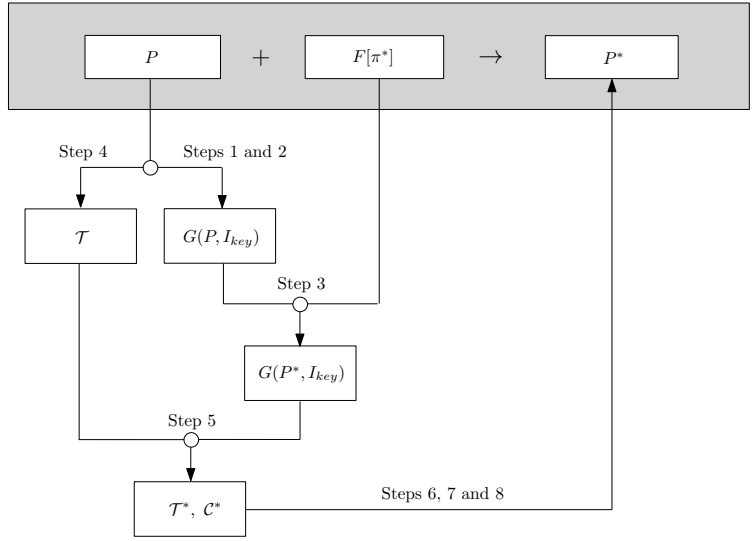}
    \medskip\medskip
    \hrule
    \centering\medskip\medskip
    \caption{\small{A block diagram of the main operations of the embedding algorithm.}}
    \label{diagram}
\end{figure}

\vspace*{0.2in}
\subsection{Embedding an RPG into a Code}
\noindent Let us now present our model's algorithm which efficiently watermarks an application program $P$ by embedding the reducible permutation graph $F[\pi^*]$ into $P$. The proposed embedding algorithm, which we call ${\tt Encode\_RPG.to.CODE}$, is described below.

\newpage

\vskip 0.0in 
\vspace*{0.15in}
\noindent Embedding Algorithm ${\tt Encode\_RPG.to.CODE}$
\begin{enumerate}
\item[1.\,]   
Take as input the source code of the program $P$, select an input $I_{key}$, and construct the call-graph $G(P, I_{key})$; let $S=(f_{main},f_s=f_{n+1}, f_n, \ldots, f_{1}, f_0=f_{t})$ be the execution sequence of the functions of call-graph $G(P, I_{key})$, that is, $f_i$ appears before $f_j$ in $S$ if $f_i$ is executed before $f_j$ with input $I_{key}$, and let $s=u_{n+1}, u_n, \ldots, u_1, u_{0}=t$ be the $n+2$ nodes of the watermark graph $F[\pi^*]$;
\vspace*{0.04in}
\item[2.\,]   
Remove the node $f_{main}$ from $G(P, I_{key})$ and assign an exact pairing (i.e., 1-1 correspondence) of the $n+2$ nodes of $G(P, I_{key})$ or, equivalently, the $n+2$ functions $f_s=f_{n+1}, f_n, \ldots, f_{1}, f_0=f_{t}$ to the nodes of $F[\pi^*]$;
\vspace*{0.04in}
\item[3.\,]   
Construct the graph $G(P^*, I_{key})$ as follows:
\begin{itemize}
  \item[3.1.\,] $V(G(P^*, I_{key}))=V(G(P, I_{key}))$, i.e., $G(P^*, I_{key})$ has the same nodes as the call-graph $G(P, I_{key})$;
  \item[3.2.\,] $E(G(P^*, I_{key}))=E(F[\pi^*])$, i.e., $(f_i, f_j)$ is an edge in $E(G(P^*, I_{key}))$ iff the corresponding $(u_i, u_j)$ is an edge in graph $F[\pi^*]$;
\end{itemize}

\vspace*{0.04in}
\item[4.\,]   
Create a call-table $\mathcal{T}$ of size $m$ which contains all the $m$ function calls $(f_i, f_j)$ as they appear in the execution trace of program $P$ with input $I_{key}$;
\vspace*{0.04in}
\item[5.\,]   
Create the tables $\mathcal{T^*}$ and $\mathcal{C^*}$, both of size $m^*$, as follows:
\begin{enumerate}
\item[5.1.\,] For each function call $(f_i, f_j)$ of table $\mathcal{T}$ do the following:
    \begin{itemize}
        \item[$\bullet$] if $(f_i,f_j)$ is a function call of $G(P^*, I_{key})$ then insert $(f_i,f_j)$ in table $\mathcal{T^*}$ and its characterization in table $\mathcal{C^*}$; in this step, $(f_i,f_j)$ is characterized as either real-forward or real-backward;
        \item[$\bullet$] if $(f_i,f_j)$ is not a function call in graph $G(P^*, I_{key})$ then:

            \begin{itemize}
                \item[$\circ$] compute the shortest path $(f_i,$ $f_{k_1},$ $f_{k_2},$ $\ldots,$ $f_{k_\ell},$ $f_j)$ from node $f_i$ to node $f_j$ in $G(P^*, I_{key})$,
                \item[$\circ$] insert the calls $(f_i, f_{k_1})$, $(f_{k_1}, f_{k_2})$, $\ldots$, $(f_{k_\ell}, f_j)$ in table $\mathcal{T^*}$, in that order, and their characterizations in table $\mathcal{C^*}$; in this step, a call is characterized as either real-forward, real-backward, water-forward, or water-backward
                    (see, Subsection~\ref{Model Components}), and
                \item[$\circ$] mark the first and last function calls of the shortest path, i.e., $(f_i, f_{k_1})$ and $(f_{k_\ell}, f_j)$, as \emph{first} and \emph{last}, respectively;
            \end{itemize}

      \end{itemize}

\item[5.2.\,] For each function call $(f_i,f_j)$ of graph $G(P^*, I_{key})$ check whether it appears in table $\mathcal{T^*}$; if not, do the following:
    \begin{itemize}
      \item[$\bullet$] find the first appearance of a function call of type $(f_o,f_i)$ in table $\mathcal{T^*}$, where $f_o$ is any function;
      \item[$\bullet$] insert the function call $(f_i, f_j)$ in table $\mathcal{T^*}$ after the call $(f_o,f_i)$ and its characterization in the corresponding row in table $\mathcal{C^*}$;
    \end{itemize}
\end{enumerate}

\vspace*{0.04in}
\item[6.\,]  Take each function call $(f_i, f_j)$ of the table $\mathcal{T^*}$ and modify the functions $f_i$ and $f_j$ of program $P$ as follows:

\begin{enumerate}
\item[6.1.\,] Add/replace call statements and locate appropriate call points in function $f_i$ as follows:

    \begin{itemize}
        \item[$\bullet$] if $(f_i, f_j)$ is a real function call then find the statement ${\tt call}$($f_j$) in function $f_i$ and locate its call-point;
        \item[$\bullet$] if $(f_i, f_j)$ is a water function call then:

            \begin{itemize}
                \item[$\circ$] if it is the first function call of a short path, then find the \emph{last} function call of that path in table $\mathcal{T^*}$, say, $(f_{k_\ell}, f_{last})$, replace the statement ${\tt call}$($f_{last}$) in function $f_i$ with the statement ${\tt call}$($f_j$), and locate its call-point;
                \item[$\circ$] otherwise, if ${\tt call}$($f_j$) does not exist in f-block of function $f_i$, add the statement ${\tt call}$($f_j$) in f$|$b-block and locate its call-point;
            \end{itemize}

    \end{itemize}

\item[6.2.\,] Insert either statement $x=x+h()$ or $x=x+g()$ before the statement ${\tt call}$($f_j$), if $(f_i, f_j)$ is characterized either as forward or backward, respectively;

\item[6.3.\,] Include both statements $x=x+h()$ or $x=x+g()$ and ${\tt call}$($f_j$) in a control statement and evaluate it as true or false using the \emph{cf-variable} $x$;

\item[6.4.\,] Include all the statements of the b-block of function $f_i$ in a control statement and evaluate it using the \emph{cf-variable} $x$;

\item[6.5.\,] Add the statement $x=x+c()$ before the f$|$b-block of function $f_j$;

\end{enumerate}

\vspace*{0.04in}
\item[7.\,]   
For each function call $(f_{out},f_j)$ of program $P$ s.t. $f_{out} \notin G(P^*, I_{key})$ and $f_j \in G(P^*, I_{key})$ do the following:

    \begin{enumerate}
        \item[7.1.\,] Find the first appearance of a function call of type $(f_o,f_j)$ in table $\mathcal{T^*}$, such that $f_o$ is any function and $f_j$ is either a real function call or the last function of a shortest path $(f_i, f_{k_1}, f_{k_2}, \ldots, f_{k_\ell}, f_j)$, i.e., $(f_o,f_j)=(f_{k_\ell}, f_j)$;

        \item[7.2.\,] Take the value of the \emph{cf-variable} $x$ of function $f_{k_\ell}$, say, ``value", and insert the statement $x = ``value"$ in function $f_{out}$ before the call-point of the statement ${\tt call}$($f_j$);
    \end{enumerate}

\vspace*{0.04in}
\item[8.\,]   
Return the source code of the modified program $P$ which is the watermarked program $P^*$;
\end{enumerate}

\vskip 0.1in 
\noindent {\bf Remark~3.4.} In Step~5 of the embedding algorithm, the edges $(f_i, f_j)$ are included into the table $\mathcal{T}$ in a specific order. This order is determined by the order they appeared in the execution trace of program $P$ with input $I_{key}$, i.e., if the function call $(f_i, f_j)$ appears before $(f_k, f_\ell)$ in the execution trace of $P$, then the edge $(f_i, f_j)$ appears before the edge $(f_k, f_\ell)$ in table $\mathcal{T}$.

\vspace*{0.1in}
\noindent {\bf Remark~3.5.} Let $(f_i, f_j)$ be an edge which is handled in Step~6 of the embedding algorithm and let the statement ${\tt call}$($f_j$) appear more that once in function $f_i$. We point out that in this case we insert both the cf-variable and cf-statement before the call-site of each statement ${\tt call}$($f_j$) in function $f_i$.

\vskip 0.1in 
\noindent {\bf The Algorithm by an Example.} In order to illustrate the working of the embedding algorithm ${\tt Encode\_RPG.to.CODE}$, we present a simple example and show the main operations (i.e., function calls) and the values of the main variables during the algorithm's execution.

\begin{figure}[t!]
    \centering
    \hrule
        \medskip\medskip
        \includegraphics[scale=0.7]{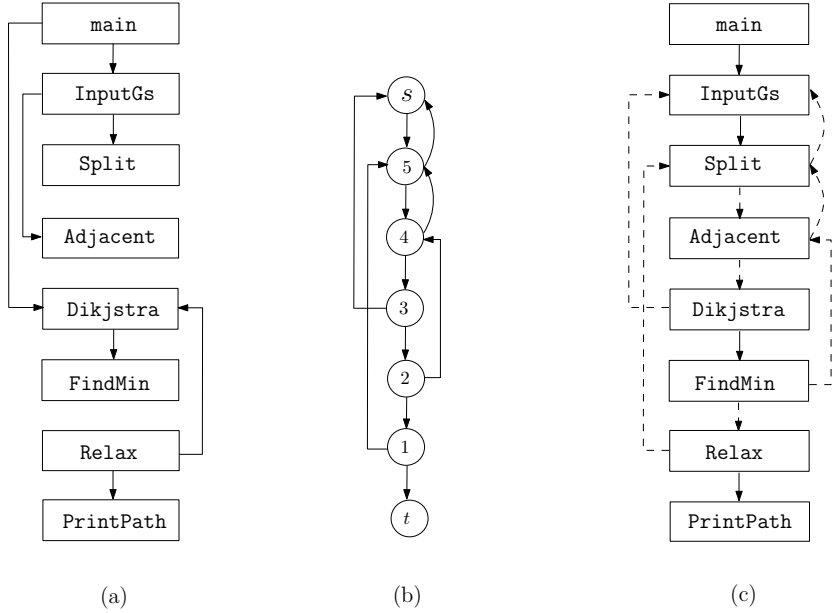}
    \medskip\medskip
    \hrule
    \centering\medskip\medskip
    \caption{\small{(a) The dynamic call-graph $G({\tt Shortest\_Path}, I_{key})$. (b) The reducible permutation graph $F[\pi^*]$ which encodes the watermark $w=2$, where $\pi^*=(3,5,1,4,2)$. (c) The dynamic call-graph $G({\tt Shortest\_Path^*}, I_{key})$.}}
    \label{Example-call-graphs}
\end{figure}

In our example, we chose the original program $P$ to be one that computes the shortest paths in a weighted graph $G$ with non-negative edge-values; it takes as input a graph $G$ and a node $s$ and computes the shortest paths from $s$ to every other node $v \in V(G)-\{s\}$. The program $P$, which we call ${\tt Shortest\_Path}$, consists of 8 functions (i.e., 7 functions plus the main) and have the property that its dynamic call-graph $G({\tt Shortest\_Path}, I_{key})$ is the same for every input $I_{key}$; see, its dynamic call-graph in Figure~\ref{Example-call-graphs}(a). Moreover, we chose to embed the watermark number $w=2$ into the source code of program ${\tt Shortest\_Path}$. The reducible permutation graph $F[\pi^*]$ which encodes the watermark $w=2$ consists of 7 nodes and is depicted in Figure~\ref{Example-call-graphs}(b). We point out that, according to our model's rules, the number $w=2$ is encoded by the SiP $\pi^*=(3,5,1,4,2)$ and it can be successfully embedded into ${\tt Shortest\_Path}$ since our program consists of 8 functions and the graph $F[\pi^*]$ contains 7 nodes; note that, the number $w=3$ is the max number which can be embedded into our program since the graph $F[\pi^*]$ for encoding the watermark $w=4$ has to contain 9 nodes (see, Section~\ref{Encode-Numbers} and also Figure~\ref{components}). The dynamic call-graph $G({\tt Shortest\_Path^*}, I_{key})$ of our watermarked program is presented in Figure~\ref{Example-call-graphs}(c). Observe that, $G({\tt Shortest\_Path^*}, I_{key})$ is isomorphic to the watermark graph $F[(3,5,1,4,2)]$.

In Figure~\ref{tables} we show the call-tables $\mathcal{T}$ and $\mathcal{T^*}$ of the programs ${\tt Shortest\_Path}$ and ${\tt Shortest\_Path^*}$, respectively, the edge characterization of each function call $(f_i, f_j)$ of the call-table $\mathcal{T^*}$ (see, Table~$\mathcal{C^*}$), and the increment of the value of the $cf$-variable $x$ in each case. More precisely, in Table $\mathcal{C^*}$ each $(f_i, f_j)$ is characterized as either $rf$ (real-forward), $rb$ (real-backward), $wf$ (water-forward), or $wb$ (water-backward), while in the case where a function call $(f_i, f_j)$ is replaced by a path of calls (shortest path) we characterize as \emph{first} (resp. \emph{last}) the first (resp. last) function call of that path. The fourth table of Figure~\ref{tables} shows the values (in parentheses) of the $cf$-variable $x$ in both functions $f_i$ and $f_j$ of the program ${\tt Shortest\_Path^*}$. Recall that according to the f-call and b-call patterns (see, Section~\ref{Model Components}), if $(f_i, f_j)$ is a forward edge we add the statement $x=x+h()$ in function $f_i$, while if $(f_i, f_j)$ is a backward edge we add the statement $x=x+g()$ in function $f_i$; in both cases we unconditionally add the statement $x=x+c()$ in function $f_j$.

In our example, we initialize the $cf$-variable $x=0$ and consider for simplicity reasons constant values for the functions $h()$, $g()$, and $c()$, that is, $h()=3$, $g()=2$, and $c()=1$. Based on the above, we take the first function call $({\tt main}, {\tt InputGs})$ of Table~$\mathcal{T^*}$ and, since it is characterized as $rf$ in Table~$\mathcal{C^*}$, we increase by $h()$ the value of the $cf$-variable $x$ in function ${\tt main}$ before the call site of ${\tt InputGs}$, i.e., we set $x=x+h()$ and thus $x = 3$. In the callee function ${\tt InputGs}$ we always increase by $c()$ the value of the variable $x$, i.e., we set $x=x+c()$ and thus $x = 4$. We observe that, the function call $({\tt Split}, {\tt InputGs})$ of Table~$\mathcal{T^*}$ is characterized as $wb$ (water-backward) and, thus, we increase by $g()=2$ the value of the variable $x$ in function ${\tt Split}$, again before the call site of ${\tt InputGs}$, i.e., $x$ becomes equal to 10 because in the previous function call $x$'s value was 8 (see, last table of Figure~\ref{tables}). Note that, by construction the shortest paths of function calls do not intersect.

\begin{figure}[t!]
\hrule
\medskip\medskip
\begin{center}
\begin{tabular}{ c c c c c}
\begin{tabular}{| l |}
\hline
{Call-table $\mathcal{T}$ }    \\
\hline \hline
{${\tt main}$ $\rightarrow$ ${\tt InputGs}$}             \\
{${\tt InputGs}$ $\rightarrow$ ${\tt Split}$}            \\
{${\tt InputGs}$ $\rightarrow$ ${\tt Split}$}            \\
{${\tt InputGs}$ $\rightarrow$ ${\tt Adjacent}$}        \\
{${\tt main}$ $\rightarrow$ ${\tt Dikjstra}$}               \\
{${\tt Dikjstra}$ $\rightarrow$ ${\tt FindMin}$}            \\
{${\tt Relax}$ $\rightarrow$ ${\tt Dikjstra}$}              \\
{${\tt Relax}$ $\rightarrow$ ${\tt PrintPath}$}             \\
{}\\
{}\\
{}\\
{}\\
{}\\
{}\\
{}\\
{}\\
{}\\
{}\\
\hline
\end{tabular}

\begin{tabular}{ c }
{ }    \\
\end{tabular}

\begin{tabular}{| l |}

\hline
{Call-table $\mathcal{T^*}$ }    \\
\hline \hline
{${\tt main}$ $\rightarrow$ ${\tt InputGs}$}             \\
{${\tt InputGs}$ $\rightarrow$ ${\tt Split}$}            \\
{${\tt Split}$ $\rightarrow$ ${\tt InputGs}$}            \\
{${\tt InputGs}$ $\rightarrow$ ${\tt Split}$}            \\
{${\tt InputGs}$ $\rightarrow$ ${\tt Split}$}            \\
{${\tt Split}$ $\rightarrow$ ${\tt Adjacent}$}             \\
{${\tt Adjacent}$ $\rightarrow$ ${\tt Split}$}             \\
{${\tt main}$ $\rightarrow$ ${\tt InputGs}$}             \\
{${\tt InputGs}$ $\rightarrow$ ${\tt Split}$}            \\
{${\tt Split}$ $\rightarrow$ ${\tt Adjacent}$}             \\
{${\tt Adjacent}$ $\rightarrow$ ${\tt Dikjstra}$}          \\
{${\tt Dikjstra}$ $\rightarrow$ ${\tt InputG}$}         \\
{${\tt Dikjstra}$ $\rightarrow$ ${\tt FindMin}$}            \\
{${\tt FindMin}$ $\rightarrow$ ${\tt Adjacent}$}           \\
{${\tt Dikjstra}$ $\rightarrow$ ${\tt FindMin}$}            \\
{${\tt FindMin}$ $\rightarrow$ ${\tt Relax}$}               \\
{${\tt Relax}$ $\rightarrow$ ${\tt Split}$}                 \\
{${\tt Relax}$ $\rightarrow$ ${\tt PrintPath}$}             \\

\hline
\end{tabular}

\begin{tabular}{ c }
{ }    \\
\end{tabular}

\begin{tabular}{| l |}
\hline
{Table $\mathcal{C^*}$ }    \\
\hline \hline
{\it rf}            \\
{\it rf}            \\
{\it wb}            \\
{\it rf}            \\
{{\it wf} -- first}      \\
{{\it wf} -- last}       \\
{\it wb}            \\
{{\it wf} -- first}      \\
{\it wf}            \\
{\it wf}            \\
{{\it wf} -- last}       \\
{\it wb}            \\
{\it rf}            \\
{\it wb}            \\
{{\it wf} -- first}      \\
{{\it wf} -- last}       \\
{\it wb}            \\
{\it rf}            \\
\hline
\end{tabular}

\begin{tabular}{ c }
{ }    \\
\end{tabular}

\begin{tabular}{| l |}
\hline
{Values of the {\it cf-variable}}    \\
\hline \hline
{$x$+=3 \ \ \ \ (3)   \ $\rightarrow$    \ $x$+=1 \ \ \ (4)}          \\
{$x$+=3 \ \ \ \ (7)   \ $\rightarrow$    \ $x$+=1 \ \ \ (8)}          \\
{$x$+=2 \ \ (10)   \  $\rightarrow$      \ \ ... \ \ \ \ \ (11)}        \\
{$x$+=3 \ \ (14)   \  $\rightarrow$      \ \ ... \ \ \ \ \ (15)}        \\
{$x$+=3 \ \ (18)   \  $\rightarrow$      \ \ ... \ \ \ \ \ (19)}        \\
{$x$+=3 \ \ (22)   \  $\rightarrow$      \ \ ... \ \ \ \ \ (23)}        \\
{$x$+=2 \ \ (25)   \  $\rightarrow$      \ \ ... \ \ \ \ \ (26)}        \\
{$x$+=3 \ \ (29)   \  $\rightarrow$      \ \ ... \ \ \ \ \ (30)}        \\
{$x$+=3 \ \ (33)   \  $\rightarrow$      \ \ ... \ \ \ \ \ (34)}        \\
{$x$+=3 \ \ (37)   \  $\rightarrow$      \ \ ... \ \ \ \ \ (38)}        \\
{$x$+=3 \ \ (41)   \  $\rightarrow$      \ \ ... \ \ \ \ \ (42)}        \\
{$x$+=2 \ \ (44)   \  $\rightarrow$      \ \ ... \ \ \ \ \ (45)}        \\
{$x$+=3 \ \ (48)   \  $\rightarrow$      \ \ ... \ \ \ \ \ (49)}        \\
{$x$+=2 \ \ (51)   \  $\rightarrow$      \ \ ... \ \ \ \ \ (52)}        \\
{$x$+=3 \ \ (55)   \  $\rightarrow$      \ \ ... \ \ \ \ \ (56)}        \\
{$x$+=3 \ \ (59)   \  $\rightarrow$      \ \ ... \ \ \ \ \ (60)}        \\
{$x$+=2 \ \ (62)   \  $\rightarrow$      \ \ ... \ \ \ \ \ (63)}        \\
{$x$+=3 \ \ (66)   \  $\rightarrow$      \ $x$+=1 \ (67)}        \\
\hline
\end{tabular}

\end{tabular}
\medskip\medskip
\hrule
\centering\medskip\medskip
\caption{The call-tables $\mathcal{T}$ and $\mathcal{T^*}$ of the programs $Shortest\_Path$ and $Shortest\_Path^*$, respectively, the edge characterization table $\mathcal{C^*}$, and the values of the $cf$-variable.}
\label{tables}
\end{center}
\end{figure}

\vskip 0.2in 
\subsection{Extracting the RPG from the Code}
We next present our WaterRPG model's algorithm for extracting the graph $F[\pi^*]$ from the program $P^*$ watermarked by the embedding algorithm ${\tt Encode\_RPG.to.CODE}$. The proposed extracting algorithm works as follows:

\vskip 0.0in 
\vspace*{0.15in} \noindent Extracting Algorithm ${\tt Decode\_CODE.to.RPG}$
\begin{enumerate}
\item[1.\,]   
Take as input the program $P^*$ watermarked by the embedding algorithm ${\tt Encode\_RPG.to.CODE}$ and run it with input $I_{key}$;
\vspace*{0.04in}
\item[2.\,]   
Construct the call-table $\mathcal{T}$ using the execution trace of the program $P^*$ with input $I_{key}$;
\vspace*{0.04in}
\item[3.\,]   
Construct the dynamic call-graph $G(P^*, I_{key})$ using the call-table $\mathcal{T}$ as follows:
\begin{itemize}
    \item[3.1.\,] take all the function calls $(f_i,f_j)$ of table $\mathcal{T}$ and add both functions $f_i$ and $f_j$ in the set $V$; note that, $V$ has $n+2$ elements since $\mathcal{T}$ contains $n+2$ different functions;
    \item[3.2.\,] take all the function calls $(f_i,f_j)$  of table $\mathcal{T}$ and add the selected pairs in the set $E$; note that, $E$ contains $2n+1$ elements;
    \item[3.3.\,] assign the set $V$ to $V(G(P^*, I_{key}))$ and the set $E$ to $E(G(P^*, I_{key}))$;
\end{itemize}
\vspace*{0.04in}
\item[4.\,]   
Remove the node $f_{main}$ from the graph $G(P^*, I_{key})$; the resulting graph is a reducible permutation graph isomorphic to $F[\pi^*]$ (see, the embedding algorithm ${\tt Encode\_RPG.to.CODE}$);
\vspace*{0.04in}
\item[5.\,]   
Compute the unique Hamiltonian path HP of the graph $G(P^*, I_{key})$; let $HP=(f_{0}, f_{1}, f_{2}, \ldots, f_{n}, f_{n+1})$ be the Hamiltonian path of $G(P^*, I_{key})$;
\vspace*{0.04in}
\item[6.\,]   
Relabel the nodes of the graph $G(P^*, I_{key})$ according to their order in the HP as follows: $f_{0}=u_{n+1}$, $f_{1}=u_{n}$, $f_{2}=u_{n-1}$, $\ldots$, $f_{n}=u_{1}$, $f_{n+1}=u_{0}$; the resulting graph $G(P^*, I_{key})$ has a unique Hamiltonian path $HP=(u_{n+1}, u_{n}, u_{n-1}, \ldots, u_{1}, u_{0})$ and thus $G(P^*, I_{key})=F[\pi^*]$;
\vspace*{0.04in}
\item[7.\,]   
Return the reducible permutation graph $F[\pi^*]$;
\end{enumerate}

\vskip 0.2in

\vspace*{0.1in}
\noindent {\bf Remark~3.6.} In Step~5 of the extracting algorithm, we compute the unique Hamiltonian path of the graph $F[P^*]$. Indeed, it has been shown that the reducible permutation graph $F[\pi^*]$ has always a unique Hamiltonian path, denoted by HP$(F[\pi^*])$, and this Hamiltonian path can be found in $O(n)$ time, where $n$ is the number of nodes of $F[\pi^*]$ (author's papers). Since $F[\pi^*]$ is isomorphic to $G'(P^*, I_{key})$ we can compute the unique Hamiltonian path HP of the graph $F[P^*]$ within the same time complexity.

\section{Implementation}

\vspace*{0.1in}
In this section we present in detail the watermarking process performed by our WaterRPG model on a Java application program $P$. We show the implementation of our watermarking process using a real  program with market-name ${\tt Laser}$ which we have downloaded from the website ${\tt www.java-gaming.org}$ containing various and different in characteristics game application programs.

Let $f_s=f_{n+1}, f_n, \ldots, f_{1}, f_0=f_t$ be the functions of the dynamic call-graph $G(P, I_{key})$, where $P={\tt Laser}$.
Recall that the functions $f_s=f_{n+1}, f_n, \ldots, f_{1}, f_0=f_t$ are
into 1-1 correspondence with the nodes $s=u_{n+1}, u_n, \ldots, u_1, u_{0}=t$ of the reducible permutation graph $F[\pi^*]$ which encodes the watermark number $w$.

We focus on the function $f_i={\tt up()}$ of the program ${\tt Laser}$; in our implementation, the important part of the Java code of $f_i={\tt up()}$ is the following:

\vspace*{0.15in}

${\tt public \ void \ up\{}$\\
\Mytab\Mytab\Mytab\Mytab ${\tt if \ (b[cx+1][cy-1-1].bgr() \ \ldots)\{}$ \\
\Mytab\Mytab\Mytab\Mytab\Mytab\Mytab\Mytab\Mytab ${\tt hlth--;}$\\
\Mytab\Mytab\Mytab\Mytab ${\tt \}}$ \\
\Mytab\Mytab\Mytab\Mytab ${\tt b[cx+1][cy].bgr(black);}$\\
\Mytab\Mytab\Mytab\Mytab ${\tt \vdots}$

\vspace*{0.1in}
\noindent We first show the straightforward case of the watermarking process on function $f_i={\tt up()}$ and, then, we proceed with advanced cases. In all cases our model uses the cf-variable $x$ which increases its value by $h()=3$, $g()=2$, and $c()=1$; see, Call Patterns in Section~3.2.

Before we proceed to watermark the function $f_i={\tt up()}$, we divide the callee functions of $f_i$ into the following three categories:

\begin{itemize}[leftmargin=14.3mm, labelindent=8.0mm, labelsep=4.0mm]
  \item[$\mathcal{A}_{callee}$\,:] contains the callee functions $f_j^1$ and $f_j^2$ of $f_i$ which correspond to forward node $u_j^1$ and
  backward node $u_j^2$ of graph $F[\pi^*]$, respectively; that is, both $f_j^1$ and $f_j^2$ are functions of the dynamic call-graph $G(P, I_{key})$.
  \item[$\mathcal{B}_{callee}$\,:] contains the callee functions $f_j^*$ of $f_i$ which are executed with the input $I_{key}$ except of $f_j^1$ and $f_j^2$; that is, $f_j^*$ is a function of the dynamic call-graph $G(P, I_{key})$.
  \item[$\mathcal{C}_{callee}$\,:] contains all the callee functions $f_j^{**}$ of $f_i$ which are not executed with the input $I_{key}$.
\end{itemize}

\vspace*{0.2in}
\noindent {\bf Naive-case Implementation}

\vspace*{0.1in}
\noindent Let $u_i$ be the node of graph $F[\pi^*]$ which corresponds to $f_i={\tt up()}$, and let $u_j^1$ and $u_j^2$ be the two nodes of $F[\pi^*]$ such that $(u_i,u_j^1)$ and $(u_i,u_j^2)$ are the forward and backward outgoing edges of node $u_i$, respectively. Let $f_j^1$ and $f_j^2$ be the two functions of $G(P, I_{key})$ which correspond to nodes $u_j^1$ and $u_j^2$, respectively; in our
implementation, $f_j^1={\tt down()}$ and $f_j^2={\tt health()}$.

We next describe in a step-by-step manner the modifications we make in function $f_i={\tt up()}$ according to the watermarking rules of our
WaterRPG model. The watermarking process of the naive-case implementation consists of the following phases:

\begin{itemize}
    \item[(I)] We first include the body of the function $f_i$ into a control statement holding opaque predicates of the cf-variable $x$. In our naive-case implementation, we use the statement ${\tt if}$-${\tt then}$-${\tt else}$ and add opaque predicates of the form ${\tt x==value}$; see, statement ${\tt if}$ ${\tt(x==271}$  ${\tt \&\&}$ ${\tt down==false)}$ ${\tt \{...\}}$ of Figure~\ref{fig:function-up}.

     Then, we handle the functions $f_j^1$ and $f_j^2$ of categories $\mathcal{A}_{callee}$; in particular, we locate
     the call-points of all the statements ${\tt call}(f_j^1)$ and ${\tt call}(f_j^2)$ in $f_i$, if any, and we do the following:
        \begin{itemize}
            \item[$\circ$] We form an f-block, in the case where $f_i$ contains $f_j^1$, by adding the statement $x=x+h()$ in a call-point before that of ${\tt call}(f_j^1)$ and including both $x=x+h()$ and ${\tt call}(f_j^1)$ into a control statement with opaque predicates using the cf-variable $x$; in our implementation, $f_j^1={\tt down()}$ and $h()=3$.
            \item[$\circ$] We similarly form a b-block, in the case where $f_i$ contains $f_j^2$, by adding the statement $x=x+g()$ instead of $x=x+h()$ as before; in our implementation, $f_j^2={\tt health()}$ and $g()=2$.
        \end{itemize}

    In the case where the function $f_i$ does not contain ${\tt call}(f_j^1)$ or ${\tt call}(f_j^2)$, we locate a
    call-point before that of the control statement ${\tt if}$-${\tt then}$-${\tt else}$ and we do the following:
            \begin{itemize}
            \item[$\circ$] If $f_i$ does not contain ${\tt call}(f_j^1)$, we add the statements $x=x+h()$ and ${\tt call}(f_j^1)$ in this order and, then, we include both $x=x+h()$ and ${\tt call}(f_j^1)$ into a control statement with conditions consisting of opaque predicates using the cf-variable $x$; recall that $h()=3$; see, statement ${\tt if}$ ${\tt(x==271)}$
                ${\tt \&\&}$ ${\tt down==true)}$ ${\tt \{...\}}$ of Figure~\ref{fig:function-up}.
            \item[$\circ$] If $f_i$ does not contain ${\tt call}(f_j^2)$, we add the statements $x=x+g()$ and ${\tt call}(f_j^2)$ in this order; we also include both statements into a control statement as before; see, statement ${\tt if}$ ${\tt(x==268)}$ ${\tt \{...\}}$ of
                Figure~\ref{fig:function-up}.
        \end{itemize}

    \item[(II)] In this phase, we locate a point in the beginning of the callee function $f_j^1$ (resp. $f_j^2$) of function $f_i$, add the statement $x=x+c()$ in this point and include $x=x+c()$ into a control statement with conditions consisting of opaque predicates using the cf-variable $x$; in our implementation $c()=1$; see, statement ${\tt if}$ ${\tt(x==267)}$ ${\tt \{...\}}$ of Figure~\ref{fig:function-up}.

    \item[(III)] We next handle all the functions $f_j^*$ of category $\mathcal{B}_{callee}$, that is, the callee functions of $f_i$ that are functions of the call-graph $G(P, I_{key})$ except of $f_j^1$ and $f_j^2$. For every direct call $(f_i, f_j^*)$ we compute the sequence $(f_i, f_{k_1}, \ldots, f_j^*)$ which corresponds to the shortest path $(u_i, u_{k_1}, \ldots, u_j^*)$ from $u_i$ to $u_j^*$ in graph $F[\pi^*]$; then, we remove the statement ${\tt call}(f_j^*)$ from $f_i$ and add either the statements $x=x+h()$ and ${\tt call}(f_j^1)$ if $(u_i, u_{k_1})$ is a forward edge or the statements $x=x+g()$ and ${\tt call}(f_j^1)$ if $(u_i, u_{k_1})$ is a backward edge in $F[\pi^*]$; in any case, we include the added statements into a control statement with conditions consisting of opaque predicates using the cf-variable $x$.

    \item[(IV)] In the last phase we handle all the functions $f_{out}$ of program $P$ which call functions that correspond to nodes of graph $F[\pi^*]$, i.e., $f_{out}$ is not a function of the call-graph $G(P, I_{key})$ and calls a function $f_j$ of $f_s=f_{n+1}, f_n, \ldots, f_{1}, f_0=f_t$. We find the first appearance of a function call of type $(f_{out},f_j)$ in table $\mathcal{T^*}$ such that $f_j$ is either a real function call or the last function of a shortest path $(f_i, f_{k_1}, f_{k_2}, \ldots, f_{k_\ell}, f_j)$ (see, Step~7 of embedding algorithm), take the value of the cf-variable $x$ of function $f_{k_\ell}$, say, ``value", and insert the statement $x = ``value"$ in function $f_{out}$ before the call-point of the statement ${\tt call}$($f_j$).
\end{itemize}

\noindent All the functions $f_j^{**}$ of category $\mathcal{C}_{callee}$ are ignored during the process of watermarking the
function $f_i={\tt up()}$ since they are not executed with the input $I_{key}$.

\begin{figure*}[!t]
\hrule\medskip
{\small The Naive-case Implementation} \hspace*{4.00cm} {\small Two Stealthy-case Implementations}
\medskip\hrule\medskip
\begin{parcolumns}[nofirstindent=true,sloppy]{3}
\colchunk[1]{\small
    ${\tt public \ void \ up\{ }$\\
    \Mytab ${\tt if (x==267)\{ }$ \\
    \Mytab\Mytab\Mytab\Mytab ${\tt x=x+1; }$\\
    \Mytab ${\tt \} }$ \\
    \Mytab ${\tt if (x==268)\{ }$ \\
    \Mytab\Mytab\Mytab\Mytab ${\tt x=x+2; }$\\
    \Mytab\Mytab\Mytab\Mytab ${\tt health(); }$\\
    \Mytab ${\tt \} }$ \\
    \Mytab ${\tt if (x==271 \&\& down==true)\{ }$\\
    \Mytab\Mytab\Mytab\Mytab ${\tt x=x+3; }$\\
    \Mytab\Mytab\Mytab\Mytab ${\tt down(); }$\\
    \Mytab ${\tt \} }$ \\
    \Mytab ${\tt if (x==271 \&\& down==false)\{ }$\\
    \Mytab\Mytab\Mytab\Mytab ${\tt if (b[cx+1][cy-1-1] \ldots )\{  }$\\
    \Mytab\Mytab\Mytab\Mytab\Mytab\Mytab\Mytab\Mytab ${\tt hlth--; }$\\
    \Mytab\Mytab\Mytab\Mytab ${\tt \}  }$\\
    \Mytab\Mytab\Mytab\Mytab ${\tt b[cx+1][cy].bgr(black); }$\\
    \Mytab\Mytab\Mytab\Mytab ${\tt \vdots }$
}
\colchunk[2]{\small
    ${\tt \tt public \ void \ up\{ }$\\
    \Mytab ${\tt x=x+1; }$\\
    \Mytab ${\tt if (x==268 \&\& down==true)\{ }$ \\
    \Mytab\Mytab\Mytab\Mytab ${\tt x=x+3; }$\\
    \Mytab\Mytab\Mytab\Mytab ${\tt down(); }$\\
    \Mytab\Mytab\Mytab\Mytab ${\tt if (x==272)\{ }$ \\
    \Mytab\Mytab\Mytab\Mytab\Mytab\Mytab ${\tt x=x+2; }$\\
    \Mytab\Mytab\Mytab\Mytab\Mytab\Mytab ${\tt health(); }$\\
    \Mytab\Mytab\Mytab\Mytab ${\tt \} }$ \\
    \Mytab ${\tt \} }$ \\
    \Mytab ${\tt else\{ }$\\
    \Mytab\Mytab ${\tt if (b[cx+1][cy-1-1] \ldots }$ \\
    \Mytab\Mytab\Mytab\Mytab ${\tt \&\& x==268)\{ }$ \\
    \Mytab\Mytab\Mytab\Mytab\Mytab\Mytab ${\tt hlth--; }$\\
    \Mytab\Mytab ${\tt \} }$ \\
    \Mytab\Mytab ${\tt b[cx+1][cy].bgr(black); }$\\
    \Mytab\Mytab ${\tt \vdots }$\\
}
\colchunk[3]{\small
    ${\tt public \ void \ up\{ }$\\
    \Mytab ${\tt x=x+1; }$\\
    \Mytab ${\tt if (x==268 \&\& down==true)\{ }$ \\
    \Mytab\Mytab\Mytab\Mytab ${\tt x=x+3; }$\\
    \Mytab\Mytab\Mytab\Mytab ${\tt down(); }$\\
    \Mytab ${\tt \} }$ \\
    \Mytab ${\tt else\{ }$\\
    \Mytab\Mytab ${\tt if (b[cx+1][cy-1-1] \ldots }$ \\
    \Mytab\Mytab\Mytab\Mytab ${\tt \&\& x==268)\{ }$ \\
    \Mytab\Mytab\Mytab\Mytab\Mytab\Mytab ${\tt hlth--; }$\\
    \Mytab\Mytab\Mytab\Mytab\Mytab\Mytab ${\tt x=x+2; }$\\
    \Mytab\Mytab\Mytab\Mytab\Mytab\Mytab ${\tt health(); }$\\
    \Mytab\Mytab ${\tt \} }$ \\
    \Mytab\Mytab ${\tt b[cx+1][cy].bgr(black); }$\\
    \Mytab\Mytab ${\tt \vdots }$\\
}
\end{parcolumns}
\hrule\medskip\medskip
\caption{The function ${\tt up()}$ of the original program ${\tt Laser}$ watermarked with the naive approach and a stealthy
    approach; the functions ${\tt down()}$ and  ${\tt health()}$ are both water functions and belong to category $\mathcal{B}_{callee}$, i.e., both are functions of $G({\tt Laser}, I_{key})$.}
\label{fig:function-up}
\end{figure*}

\vspace*{0.1in}
\noindent {\bf Remark~4.1.} In order to avoid a large increment of the value of cf-variable $x$, in the implementation Phase (I), if a call function is included into a loop structure we could apply the call patterns as follows: (i) we add the assignment of the cf-variable (i.e., $x=x+h()$ or $x=x+g()$ assignments) outside this loop, and (ii) we add the statement $x=value\_before\_loop$ inside and before the end of the loop, where $value\_before\_loop$ is the value of $x$ before the execution of this loop.

\vspace*{0.2in}
\noindent {\bf Stealthy-case Implementation}

\vspace*{0.1in}
\noindent We next show properties and modification rules of the model's call patterns based on which we can stealthily watermark a Java application program $P$. The main modification cases, which we call stealthy cases, supported by the WaterRPG model are the following:

\begin{itemize}
    \item [(S.1)] {\it Making nested patterns}: We can merge f-statements and b-statements in any way; for example,
    we can include the control b-statement ${\tt if}$ ${\tt(x==268)}$ ${\tt \{...\}}$ inside the
    f-statement ${\tt if}$ ${\tt(x==271}$ ${\tt \&\&}$ ${\tt down==true)}$ ${\tt \{...\}}$ after the statement ${\tt call}(f_j^1)={\tt up()}$; we
    appropriately change their opaque predicates; see, Figure~\ref{fig:function-up}.

    \item [(S.2)] {\it Adding multiple water-calls}: Since water-calls do not affect the functionality of the program, we can
    add multiple water-calls in the body of the function $f_i={\tt up()}$. Our aim is to increase the complexity of the source
    code making thus difficult for an attacker to understand it, the more the complexity the more the extend of the code.

    \item [(S.3)] {\it Removing control statements}: We can remove the control statement that includes the statement $x=x+c()$
    of a function $f_i={\tt up()}$ (Phase III); we can do that in the case where $f_i$ is called by a function of category $\mathcal{C}_{callee}$; note
    that, functions of category $\mathcal{C}_{callee}$ do not modify the value of the cf-variable $x$.

    \item [(S.4)] {\it Constructing complex opaque predicates}: We can construct more complex opaque predicates thus making
    the control flow of a program more difficult for an attacker to analyze it. In Phase~I, we added opaque predicates of
    the form $({\tt x==value_1}$ $||$ ${\tt x==value_2}$ $||$ $\ldots$ $||$ ${\tt x==value_m})$, whereas in the stealthy case we
    evaluate the cf-variable in a range of values $({\tt x<=value_i}$ $\&\&$ ${\tt x>=value_j})$ by adding logical and relational
    operators.

    \item [(S.5)] {\it Merging control statements}: We can merge control statements that we added in program $P^*$ with
    program's original control statements by appropriately merging their corresponding logical expressions.

    \item [(S.6)] {\it Assigning complex expressions}: In the naive case the incremental functions of statements $x=x+h()$
    and $x=x+g()$ have constant values $h()=3$ and $g()=2$, respectively. We can easily use any complex function for $h()$
    and $g()$ in order to systematically increase the cf-variable $x$.

    \item [(S.7)] {\it Using more cf-variables}: We can use more that one cf-variable to control the flow of the watermarked
    program $P^*$. We built relationships between the cf-variables in order to be used interchangeably throughout the
    execution phase. We establish thresholds that determine the use of different cf-variables.
\end{itemize}

\noindent In Figure~\ref{fig:function-up} we present two stealthy-case implementations of the function ${\tt up()}$ of the original program ${\tt Laser}$ (second and third code). The functions ${\tt down()}$ and ${\tt health()}$ are both water functions and belong to category $\mathcal{B}_{callee}$, that is, both are functions of $G({\tt Laser}, I_{key})$.

\vskip 0.5in 
\section{Model Evaluation}

\noindent Having designed a static or dynamic software watermarking model, it is very important to evaluate it under various criteria in order to gain information about its practical behavior. Several criteria have been appeared in the literature and used for evaluating the properties of a proposed watermarking model and showing its strong and weak implementation points \cite{CHCTS09}. It is a common belief that a good watermarking model must have at least the following characteristics \cite{CCKT03}:

\begin{itemize}
  \item[$\circ$] a software watermarking model should not adversely affect the size and execution time of the program $P$;
  \item[$\circ$] the ratio of the number of bits of the whole program $P$ to the number of bits encoded by the watermark $w$ should be high;
  \item[$\circ$] a model must be resilient against a reasonable set of malicious watermarking attacks;
  \item[$\circ$] both host program $P$ and watermarked program $P^*$ should have similar statistical properties.
\end{itemize}

\noindent In our work, we use various criteria which mainly aim to evaluate our WaterRPG model's performance and its resiliency. More precisely, we propose a set of evaluation criteria consisting of two main categories:

\begin{itemize}
    \item[(I)] Performance criteria.
    \item[(II)] Resilience criteria.
\end{itemize}

\noindent The Performance criteria (or, P-criteria) concern the behavior of the resulting watermarked program $P^*$ and the quality and effectiveness of the embedded watermark $w$, while the Resilience criteria (or, R-criteria) concern the robustness and resistance of the embedded watermark $w$ against malicious user attacks.

For our evaluation process, we implemented the WaterRPG model on Java application programs and experimentally evaluated it under several P-criteria and R-criteria. More precisely, we selected a number of Java application programs downloaded from the free non commercial game database website ${\tt www.java-gaming.org}$ and watermarked them using the two watermarking approaches supported by our WaterRPG model, i.e., the Naive approach, and the Stealthy approach. The selected Java programs are almost of the same size and are watermarked by embedding watermarks of three different sizes; we use watermarking graphs $F[\pi^*]$ having number of nodes $n=11$, $n=13$, and $n=15$.

All the experiments were performed on a computer with dual-core 2.0 GHZ processors, 3.0 GB of main memory under Linux operating system using Java version 1.6.0.26 of the SDK (Software Development Kit).

%
%

\vspace*{0.1in}
\subsection{Performance}
\noindent The performance criteria (or, P-criteria) mostly focus only on how much a watermarking model modifies the code of a program $P$. As these criteria range in satisfactory levels, both programs $P$ and $P^*$ have almost identical execution behavior and similar codes, and thus the code associated with the watermark $w$ is very likely to pass unnoticed by the attacker's eyes; in our classification, the P-criteria are divided into the following two main categories:

\begin{itemize}
  \item[$\bullet$\,] {\bf Data-rate}: The data-rate criterion measures the ratio $|w|/|P|$, where $|w|$ is the size of the embedded watermark $w$ and $|P|$ is the size of the original program $P$. A model should have a high data-rate so that it can embed a large message.

  \item[$\bullet$\,] {\bf Embedding overhead}: The additional execution time and space caused by embedding the watermark $w$ into program $P$, that is,
        \begin{itemize}
            \item[(i)] time overhead, and
            \item[(ii)] space overhead: (ii.a) disk space usage, and (ii.b) heap space usage.
        \end{itemize}

  \item[$\bullet$\,] {\bf Part protection}: The part protection criterion evaluates how well the watermark is distributed or spread throughout the entire code of $P$. This is an important performance property of program $P^*$ because it decreases the probability that the watermark will be altered or destroyed when small changes are made to program $P^*$.

  \item[$\bullet$\,] {\bf Credibility}: The credibility criterion evaluates how much detectable the watermark is. The embedded watermark should be easily extracted from $P^*$ and the detector (i.e., the extracting algorithm) should minimize the probability to generate false positives and false negative results.

  \item[$\bullet$\,] {\bf Stealth}: A watermarked program $P^*$ has the stealth property if the embedded watermark should exhibit the same properties as the code of $P$ or data around it and thus it should be difficult to detect. In other words, $P^*$ should have characteristics that are not different from a typical program so that an attacker can not use these characteristics to locate and attack the watermark.
\end{itemize}

\noindent Let us now discuss on the performance of the WaterRPG model and let us first focus on the data-rate of our model.

\vspace*{0.1in}
\noindent {\bf Data-rate}. This criterion essentially depends on the size of the watermark $w$ or, in our model, of the size of the embedding watermark graph $F[\pi^*]$. We consider that the size of the watermark graph is the number of vertices that it contains. In order to measure the data-rate ratio $|w|/|P|$, we compute the size of the original program $P$ by counting the number of functions it has, since in our model we assign an exact pairing of the nodes of $F[\pi^*]$ to the functions of $P$. We claim that our model has high data rate for large programs since in such programs we are able to encode a watermark graph less than or equal to programs' size. According to our model for encoding a number as reducible permutation graph a relatively large graph encodes a large set of different integer numbers.

\vspace*{0.1in}
\noindent {\bf Embedding overhead}.
In order to evaluate the embedding overhead of our WaterRPG model we choose the parameters (i) execution time, (ii.a) disk usage, and (ii.b) heap space usage. We measure these parameters on the selected Java application programs $P$ and the corresponding watermarked programs $P^*$ under both the naive and stealthy approaches. In the evaluation process, each program is executed $``n"$ times with different inputs. The run-time of each tested program is computed by taking the difference of the start-value and the end-value of the Java method ${\tt System.currentTimeMillis()}$.

The execution time overhead is proportional to the size of the watermarking graph $F[\pi^*]$. The experimental
results in Table~\ref{tab:time} indicate that for a graph $F[\pi^*]$ on $n=11$, $n=13$ and $n=15$ nodes the execution
time of the naive watermarking causes a slight increase of $5.25\%$, $7.65\%$ and $11.07\%$, respectively, while the
corresponding increments for the stealthy case are even smaller.

\begin{figure}[t!]
  \centering
    {\epsfig{file = 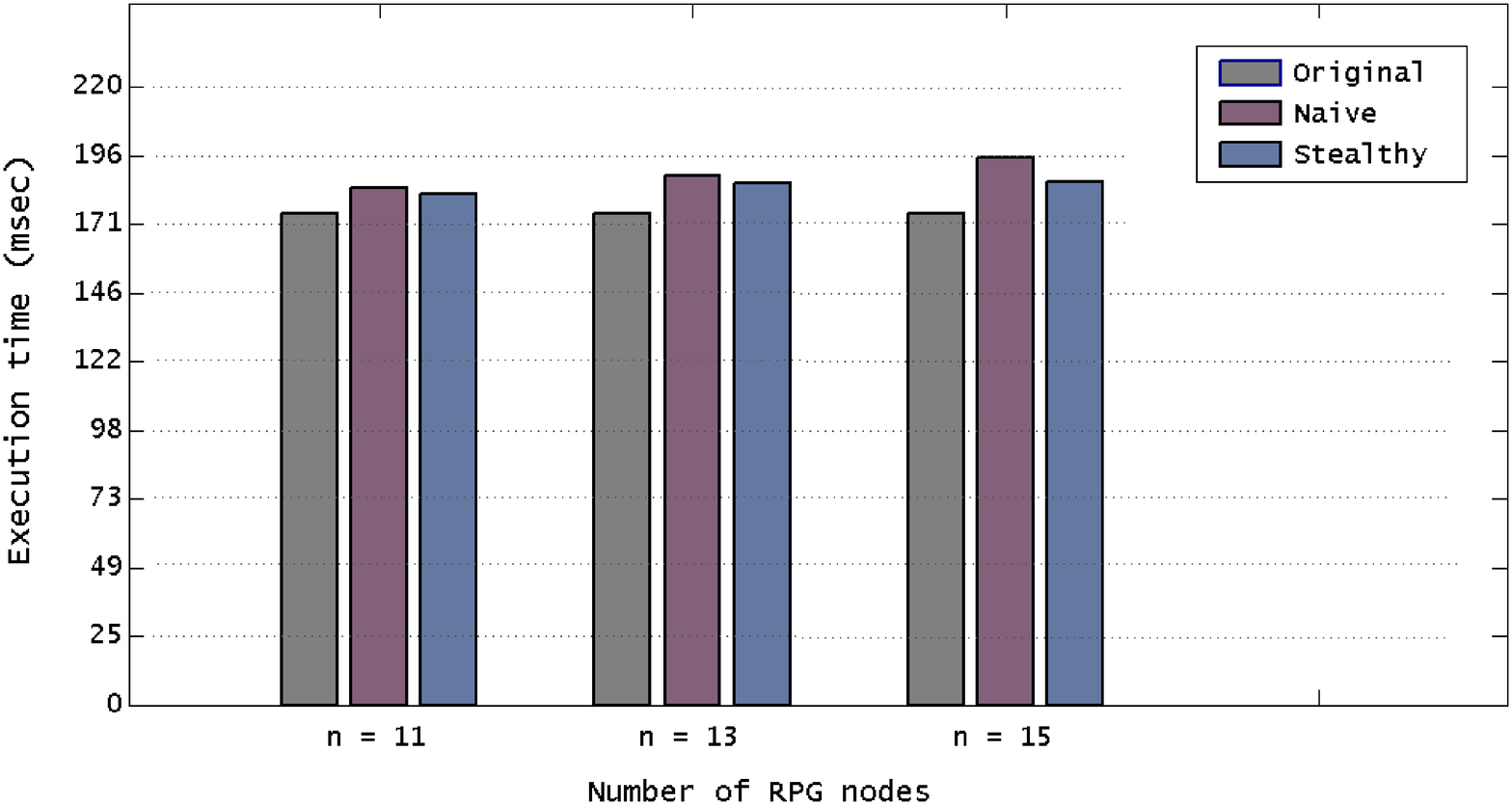, width = 11.0cm}}
  \vspace{-0.4cm}
  \vspace{0.0cm}
%
  \centering
    {\epsfig{file = 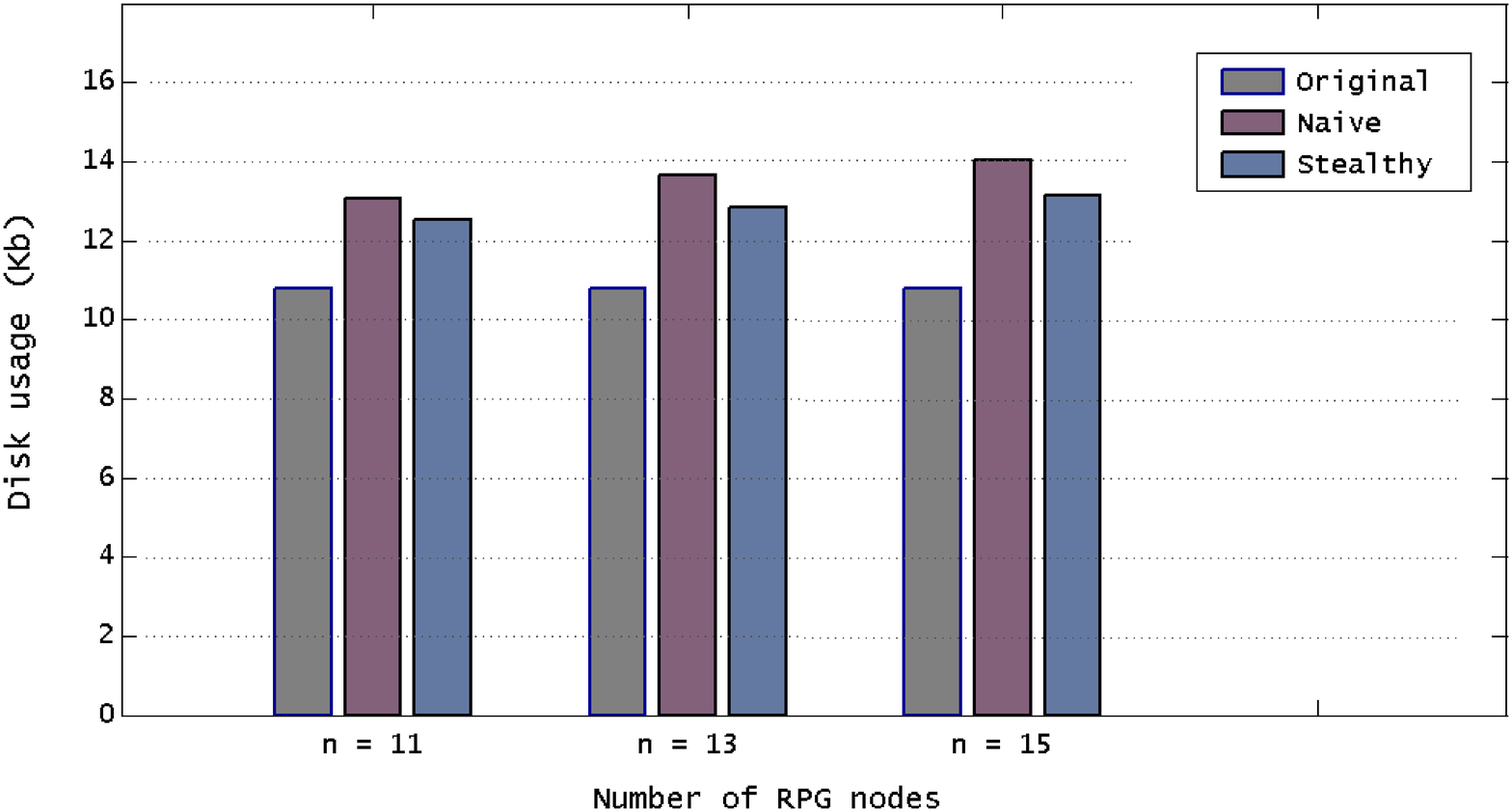, width = 11.0cm}}
  \vspace{-0.4cm}
  \vspace{0.0cm}
%
  \centering
    {\epsfig{file = 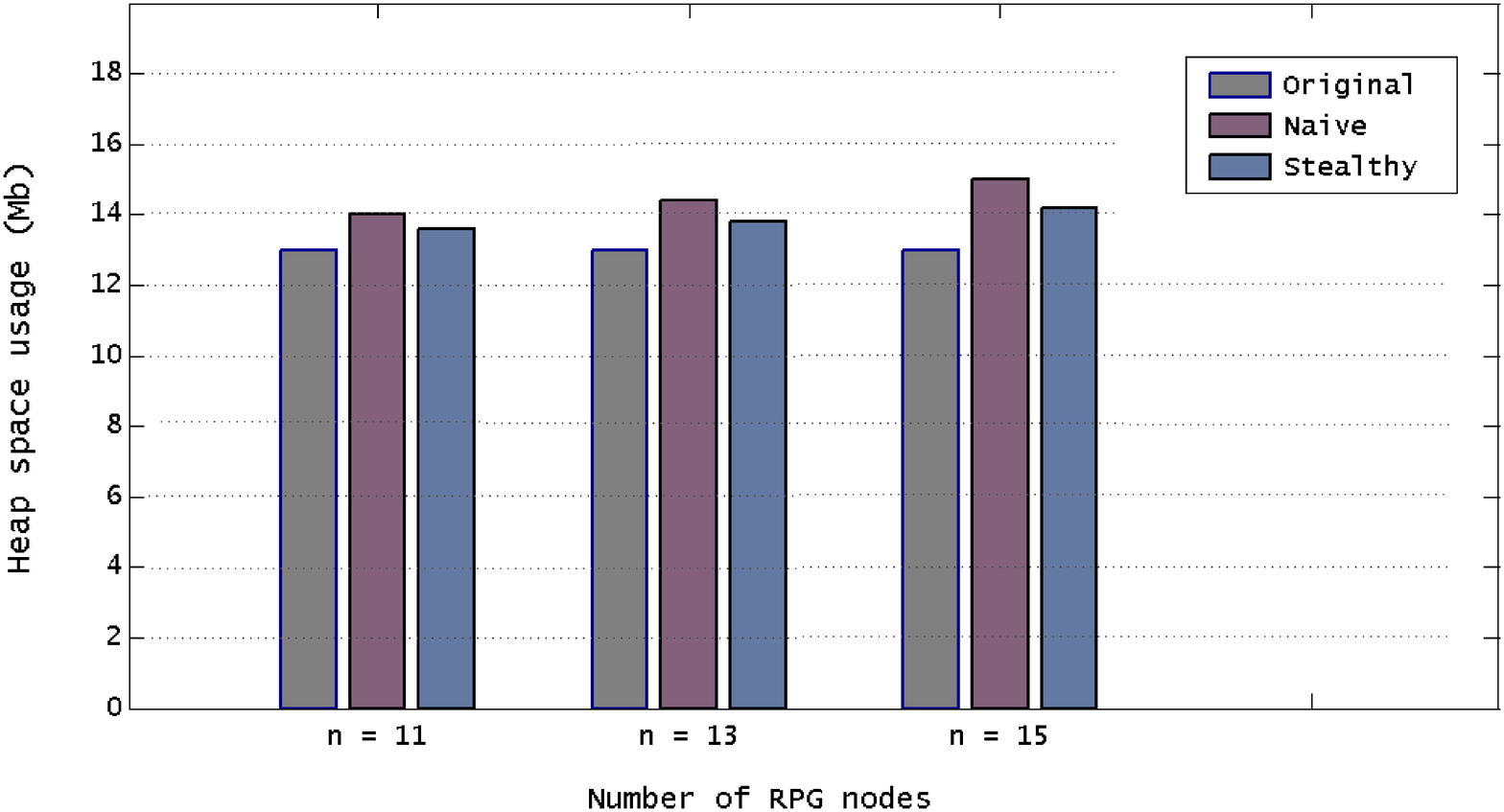, width = 11.0cm}}
  \vspace{-0.0cm}
  \caption{Graphical representation of the results for parameters (i) Execution time, (ii.a) Disk usage and (ii.b) Heap space usage of the Original program $P$, and the corresponding watermarked program under both the Naive $P^*_N$ and Stealthy $P^*_S$ approaches.}
  \vspace{0.0cm}
  \label{fig:plot}
\end{figure}

The disk storage requirements of programs $P^*$ compared to $P$ increases as the number of nodes of the graph $F[\pi^*]$ increases.
Applying the stealthy approach a noteworthy amount of storage memory is saved because many of the control statements and opaque
predicates that were not necessary to maintain proper functionality of the program $P^*$ removed safely from the
code. Table~\ref{tab:disk} illustrates the percentage increment of disk demand for $P_N^*$ and $P_S^*$, as well
as the improvement caused by the stealth approach in comparison to the naive.
The experimental results show that our WareRpg watermarking model has a similar performance for the heap space usage;
see, Table~\ref{tab:heap}. The results for all the evaluating parameters are also depicted in a graphical form in Figure~\ref{fig:plot}.

\begin{table}[t!]
\centering
\caption{Execution Time (msec)}
\vspace*{0.0in}
\begin{tabular}{|c|r|r|r|} \hline
Nodes in $F[\pi^*]$ & \phantom{x}$ P \rightarrow P_N^*$\phantom{x} & \phantom{x}$ P \rightarrow P_S^* $\phantom{x} &
\phantom{x}$ P_N^* \rightarrow P_S^*$\phantom{x} \\
\hline\hline
    11  & +5.25\%\phantom{x}  & +3.82\%\phantom{xx}  & -1.37\%\phantom{xx}\\
    \hline
    13  & +7.65\%\phantom{x}  & +5.99\%\phantom{xx}  & -1.56\%\phantom{xx}\\
    \hline
    15  & +11.07\%\phantom{x}  & +9.19\%\phantom{xx}  & -1.72\%\phantom{xx}\\
\hline
\end{tabular}
\label{tab:time}
%
\centering
\vspace*{0.2in}
\caption{Disk Usage (Kb)}
\vspace*{0.0in}
\begin{tabular}{|c|r|r|r|} \hline
Nodes in $F[\pi^*]$ & \phantom{x}$ P \rightarrow P_N^*$\phantom{x} & \phantom{x}$ P \rightarrow P_S^* $\phantom{x} &
\phantom{x}$ P_N^* \rightarrow P_S^*$\phantom{x} \\
\hline\hline
    11  & +20.98\%\phantom{x}  & +16.71\%\phantom{xx}  & -3.65\%\phantom{xx}\\
    \hline
    13  & +26.35\%\phantom{x}  & +18.81\%\phantom{xx}  & -6.34\%\phantom{xx}\\
    \hline
    15  & +30.10\%\phantom{x}  & +21.76\%\phantom{xx}  & -6.85\%\phantom{xx}\\
\hline
\end{tabular}
\label{tab:disk}
%
\centering
\vspace*{0.2in}
\caption{Heap Space Usage (Mb)}
\vspace*{0.0in}
\begin{tabular}{|c|r|r|r|} \hline
Nodes in $F[\pi^*]$ & \phantom{x}$ P \rightarrow P_N^*$\phantom{x} & \phantom{x}$ P \rightarrow P_S^* $\phantom{x} &
\phantom{x}$ P_N^* \rightarrow P_S^*$\phantom{x} \\
\hline\hline
    11      & +7.69\%\phantom{x}    & +4.61\%\phantom{xx}   & -2.94\%\phantom{xx}\\
    \hline
    13      & +10.76\%\phantom{x}   & +6.15\%\phantom{xx}   & -4.34\%\phantom{xx}\\
    \hline
    15      & +15.38\%\phantom{x}   & +9.23\%\phantom{xx}   & -5.63\%\phantom{xx}\\
\hline
\end{tabular}
\label{tab:heap}
\end{table}

Towards the evaluation of the space overhead of our watermarking method we compute the total amount of the bytecode
instructions added to watermarked program $P^*$. In particular, we compute the percentage of the increment resulted by
adding control statements, functions calls and variable assignments to the program $P$. To this end, we count the bytecode
instructions of watermarked programs $P^*_N$ and $P^*_S$ that belong to four main categories: (i) Control statements, (ii)
Invocations, (iii) Assignments, and (iv) Rest instructions; see, Table~\ref{tab:instructions}. Note that the category (iv)
contains all the bytecode instructions that remain unchanged after the watermarking process. Indicatively, in Table~\ref{tab:laser-instructions} we show the number of some bytecode instructions of the tested Java application programs $P$.

\begin{table}[b!]
\centering
\vspace*{0.06in}
\caption{Three Group of Bytecode Instructions}
\vspace*{0.0in}
\begin{tabular}{|l | r | r | r|} \hline
Bytecode\phantom{x} & \phantom{}$P$\phantom{xxx} & \phantom{} $P^*_N$\phantom{xx} & \phantom{} $P^*_S$\phantom{xx.}\\
\hline\hline
Control Statements \phantom{x.}  & \phantom{.}519.4\phantom{x.}   & \phantom{.}42.0\%\phantom{x}    & \phantom{.}25.0\%\phantom{x} \\
Invocations         & \phantom{.}188.3\phantom{x.}   & \phantom{.}10.6\%\phantom{x}    & \phantom{.}10.6\%\phantom{x} \\
Assignments         & \phantom{..}1346.7\phantom{x.}  & \phantom{..}45.5\%\phantom{x}    & \phantom{..}32.4\%\phantom{x} \\
Rest Instructions   & \phantom{.}941.6\phantom{x.}   & \phantom{.}0\%\phantom{x}       & \phantom{.}0\%\phantom{x} \\
\hline
\end{tabular}
\label{tab:instructions}
%
%
\centering
\vspace*{0.2in}
\caption{Indicative Bytecode Instructions of each Group}
\vspace*{0.0in}
\begin{tabular}{|l | r | r | r|} \hline
Bytecode\phantom{x} & \phantom{}$P$\phantom{xxx} & \phantom{} $P^*_N$\phantom{xx} & \phantom{} $P^*_S$\phantom{xx.}\\
\hline\hline
Control Statements &  &  &  \\
    \phantom{xxx}${\tt if\_icmpne}$	& 19,2\phantom{xx}    & 78,1\phantom{xx}    & 57,6\phantom{x.} \\
    \phantom{xxx}${\tt ifne}$      	& 3,1\phantom{xx}     & 4,5\phantom{xx}     & 4,5\phantom{x.} \\
	    \phantom{xxx}${\tt goto}$  	& 43,5\phantom{xx}    & 45,3\phantom{xx}    & 45,3\phantom{x.} \\

Invocations &  &  &  \\
    \phantom{xxx}${\tt invokevirtual}$    & 188,2\phantom{xx}   & 208,4\phantom{xx}   & 208,4\phantom{x.} \\

Assignments &  &  &  \\
    \phantom{xxx}${\tt iconst\_1}$ 	& 186,2\phantom{xx}   & 202,7\phantom{xx}  & 202,7\phantom{x.} \\
    \phantom{xxx}${\tt getstatic}$ 	& \phantom{xx}368,6\phantom{xx}   & \phantom{x}614,4\phantom{xx}  & \phantom{x}529,5\phantom{x.} \\
    \phantom{xxx}${\tt iadd}$     	& 84,9\phantom{xx}    & 132,8\phantom{xx}  & 132,8\phantom{x.} \\
    \phantom{xxx}${\tt alaod\_0}$  	& 136,7\phantom{xx}   & 156,2\phantom{xx}  & 156,2\phantom{x.} \\

Rest Instructions &  &  &  \\
    \phantom{xxx}${\tt dup}$       	& 33,6\phantom{xx}    & 33,6\phantom{xx}    & 33,6\phantom{x.} \\
    \phantom{xxx}${\tt ldc}$       	& 19,7\phantom{xx}    & 19,7\phantom{xx}    & 19,7\phantom{x.} \\
\hline
\end{tabular}
\label{tab:laser-instructions}
\end{table}

\vspace*{0.1in}
\noindent {\bf Part protection}. The idea behind the property of part protection is to split the watermark into pieces and then broadly spread it across the application program $P$. The splitted watermark $w$ has a better chance to survive if an attack modification on some $w$'s parts does not affect the recognition process. The more the part protection is increased, the more likely a watermark remains unchanged after a possible theft or modification of a portion of the whole watermarked code.

In our case, we do not split the watermark in order to encoded it into code, and thus our model has a low part protection. However, an attack modification on a part of our watermark code may cause incorrectness of $P^*$ unless the attacker goes through all the parts of the watermark code and makes appropriate modifications. On the other hand, since we can encode the same number $w$ into more than one reducible permutation graphs $F[\pi^*]$ (see, \cite{CN12b}), our model could obtain higher part protection by encoding multiple water-graphs $F[\pi^*]$ using different input sequences which produce different dynamic call graphs protecting thus larger code area.

\vspace*{0.1in}
\noindent {\bf Credibility}. The credibility of a watermarking model is dependent on how detectable the watermark is. Concerning our model, we should point out that the rates of false positive and false negative outcomes are noticeable low in the case where the program $P^*$ has not being attacked. In addition, even the watermarked code undergoes attacks, the execution sequence $S$ of the functions of $P^*$ is very hardly distorted. Nevertheless, an attacker could make such a distortion possible but then he has to replace the sequence of function calls $S$ with a sequence $S'$ in order to produce exactly the same output keeping the functionality and correctness of the remaining code; it is very time consuming for an attacker to find an $S'$ and test it under various inputs.

\vspace*{0.1in}
\noindent {\bf Stealth}. Some common attacks against watermarking systems begin by identifying the code composing the watermark.
To resist such attacks, watermarking should be stealthy: the watermark code embedded to a program $P$ should be locally indistinguishable from the rest code of $P$ so that it is hidden from malicious users.

The code embedded to program $P$ by our watermarking model WaterRPG is not highly unusual since our model modifies the existing source code of $P$ by only altering its control flow in order to produce, during the execution of $P^*$ with the secret input $I_{key}$, a dynamic call graph isomorphic to the watermark graph $F[\pi^*]$. More precisely, our model does not add any dead or dummy code but only encodes the graph  $F[\pi^*]$ using three groups of bytecode instructions: (i) call functions, (ii) control statements, and (iii) variable assignments. Most of these instructions are already used in the original source code $P$ and thus the embedded watermark code is quite difficult to be located in the watermarked program $P^*$. The experimental results indicate that there is an increment from 10.6\% to 45.5\% (resp. from 10.6\% to 32.4\%) of instructions of these three groups in the naive (resp. stealthy) implementation. Table~\ref{tab:instructions} shows the number of bytecode instructions, on average, of each of the three instruction groups of the tested programs $P$ and the increments of these instructions in both naive  $P^*_N$ and stealthy $P^*_S$ implementation cases, while Table~\ref{tab:laser-instructions} depicts the number of some indicative bytecode instructions of the three instruction groups.

\vspace*{0.1in}
\subsection{Resilience}
\noindent The resilience criteria (or, R-criteria) mainly focus on how the embedded watermark resists against attacks made by malicious users. These attacks are either targeted-attacks on the code composing the watermark $w$ or widespread-attacks on the whole code of program $P^*$. According to the type of attacks, the R-criteria are divided into the following two main categories:


\begin{itemize}
\item[$\bullet$\,] {\bf Water-resilience}: The water-resilience criteria measure the resistance of the watermark $w$ against attacks on its own code; we call these attacks targeted-attacks or water-attacks. In this case the attacker first detects the watermark $w$, that is, the code of program $P^*$ associated with the watermark $w$, and then makes specific operations on that code in order to
    \begin{itemize}
        \item[$\circ$] remove (e.g., by subtracting part of the watermark, or even the whole watermark),
        \item[$\circ$] destroy (e.g., by applying semantics preserving transformations so that $w$ can be undetectable, i.e., the detector can not find the watermark), or even
        \item[$\circ$] alter the watermark $w$ (e.g., by changing the structure of the embedded watermark $w$ so that it causes the extracting algorithm to produce a different watermark $w'$).
    \end{itemize}
    The attacker can also add a new watermark $w'$ into $P^*$ without modifying the existing $w$ in order to confuse the detector.

\item[$\bullet$\,] {\bf Code-resilience}: The code-resilience criteria measure the resistance of the watermark $w$ against attacks made on the whole code of the watermarked program $P^*$; we call these attacks widespread-attacks or code-attacks. In the case the attacker fails to detect the code of program $P^*$ associated with the watermark $w$, and thus he makes attacks in the whole code aiming in this way to maximize the distortion of possible watermarking protections; in our classification, the code-resilience criteria include:
    \begin{itemize}
        \item[$\circ$] obfuscation (e.g., by transferring a reducible flow graph to non-reducible),
        \item[$\circ$] optimization (e.g., by removing information for debugging with an automated tool, such as ProGuard),
        \item[$\circ$] de-compilation (e.g., by using a malicious tool, such as Java-Decompiler), and
        \item[$\circ$] language-transformation (e.g., by converting a watermarked program $P^*$ from C++ to Java).
    \end{itemize}
\end{itemize}

\noindent As in the case performance criteria, we next discuss on the resilience of our WaterRPG model focusing first on the water-resilience criteria.

\vspace*{0.1in}
\noindent {\bf Water-resilience}. The water-attacks take place when the code of the watermark $w$, the graph $F[\pi^*]$ in our model, is known to the attacker. In this case, he makes attacks on the structure of watermark graph $F[\pi^*]$ in order to destroy it or even to remove $F[\pi^*]$ from program $P^*$.

Our model embeds the watermark graph $F[\pi^*]$ into an application program $P$ by using opaque predicates in specific control statements in order to manipulate the flow of selected function calls of the watermarked program $P^*$ so that it reserves an appropriate execution, that is, $O(P,I)=O(P^*,I)$ for every input $I$. In general, it is hard for an attacker to deduce an opaque predicate at run time. Specifically, the usage of opaque predicates in our model enables us to dictate the execution flow of function calls and also makes the programs' control flow difficult for an attacker to analyze it either statically or dynamically. In fact, our model creates dependencies on the data between the original program $P$ and the watermark $F[\pi^*]$ making thus the watermark graph $F[\pi^*]$ operational part of the program $P$. This causes our model resilient to water-attacks.

Indeed, if the attacker makes a modification in a value of a cf-variable $x$ in a call-site $p$ (e.g., he increases $x$ by a constant $c$), then he has to properly modify all the values of all the cf-variables in every call-site of the execution flow after $p$ (e.g., by adding the same constant $c$) so that $P^*$ still remains functional, but then the watermark $F[\pi^*]$ also remains unchained. Moreover, if the attacker tries to remove any of the {\it cf-statements} then the program $P^*$ is not longer functional since a ``gap" is occurred on the execution flow of $P^*$ (e.g., the program executes a part of a code which in same cases produces incorrect results). What is more, every {\it cf-statement} produces different results, so our model withstands common subexpression elimination.

Additionally, if the attacker modifies either a real-call or a water-call then the program $P^*$ is no longer operational. Indeed, let the attacker modifies a water-call ${\tt call}$$(f_j)$ in function $f_i$ which appears in a path $(f_a$, $\ldots$, $f_k$, $f_i$, $f_j$, $\ldots$, $f_b)$. Then, he has to make several modifications on function calls, among which the deletion of ${\tt call}$$(f_j)$ from function $f_i$ and the replacement of ${\tt call}$$(f_i)$ with ${\tt call}$$(f_j)$ in function $f_k$, in order to remain the program $P^*$ functional. Such modification have very low probability to destroy the structure of the embedded watermark $F[\pi^*]$ without breakdown the functionality of program $P^*$.

It is worth noting that any code modification which destroys the structure of $F[\pi^*]$ can be detected with high probability by our model WaterRPG due to error-correcting properties of the watermark graph $F[\pi^*]$. Finally, we should point out that our model does not add any dead or dummy code in the watermarked program $P^*$ and also it does not use any mark during the embedding process in order to locate and extract the watermark from $P^*$.

\vspace*{0.1in}
\noindent {\bf Code-resilience}. Resiliency against code-attacks refers to the ability of a watermarking model to recognize a watermark even after the program $P^*$ has been attacked or subjected to code transformations such as translation, optimization and obfuscation \cite{MC06}. The code-attacks take place when the attacker fails to detect the code of program $P^*$ associated with our watermark $F[\pi^*]$. In this case, the attacker broadly applies code transformation attacks in the whole code of $P^*$ in order to reduce the ability of the model's recognizer to extract the watermark.

Roughly speaking, the goal of an obfuscation attack is to make a model's recognizer to hardly extract the watermark from $P^*$. Our model watermarks an application program $P$ in such a way that it withstands several obfuscation attacks among which layout-obfuscation and data-obfuscation attacks. Indeed, the experiments showed that the water-graph $F[\pi^*]$ can be efficiently extracted even the code of program $P^*$ has been subjected to control-obfuscation attacks such as expression reordering or loop reordering; some of our experiments were performed with an automated tool (e.g., ProGuard). It is fair to mention that our model does not properly operate on some other control obfuscation attacks such as aggregation including inline functions and outline functions; for example, if an attacker split a function or merge functions of $P^*$ associated with the code of the watermark $F[\pi^*]$, he actually makes a targeted attack both on vertices and edges of watermark graph $F[\pi^*]$ destroying thus the structure of $F[\pi^*]$.

As far as the optimization attacks are concerned we point out that they are mainly applied by a compiler or interpreter into the executable program altering the generated traces of program $P^*$; note that our WaterRpg model encodes execution trace watermarks. The embedded watermark $F[\pi^*]$ withstands on such attacks since any removal of function calls, register reallocation, or information for debugging does not affect the structural properties of the embedding graph $F[\pi^*]$. Again, the experimental study showed that our WaterRPG withstands on a relatively large subset of optimization attacks including semantic-preserving transformations, shrinking, and also it withstands on the most time consuming operation namely language transformation (i.e., the attacker rewrites the whole code of $P^*$ in another language).

A watermarking model must also be resilient against a reasonable set of de-compilation attacks. Thus, in our experimental study, we also included the evaluation of our watermarking model WaterRPG against de-compilation attacks. More precisely, we tested our programs with a revere engineering tool (e.g., Java Decompiler) and figure out that in all the cases that WaterRPG successfully extracts the watermarking graph $F[\pi^*]$ from the watermarked programs $P^*_N$ and $P^*_S$; indeed, in all the cases the dynamic call-graph $G(P^*, I_{key})$ taken by the input $I_{key}$ were isomorphic to graph $F[\pi^*]$.

\vskip 0.3in 
\section{Concluding Remarks}

\noindent Through the evaluation of WaterRPG, we showed that our model has zero false positive and false negative rates in the case where the watermarked code has not been attacked. Indeed, it is true because the execution of the watermarked program $P^*$ with the secret input sequence always builds a call graph $G(P^*, I_{key})$ which is isomorphic with the water-graph $F[\pi^*]$.

The execution time and space overhead varies depending on the size of the embedded watermark; in fact, the overhead increases linearly in the size of the water-graph $F[\pi^*]$. It is worth noting that the data-rate is directly correlated with the number of functions used or, equivalently, with the size of the water-graph. In the case where the code (in bits) of the original program $P$ is large enough, our model has high data-rate and extremely low embedding overhead. We point out that the number of nodes of the water-graph $F[\pi^*]$ affects the number of functions we use for embedding. Thus, it is possible to use fewer functions which would result in a graph $F[\pi^*]$ with fewer nodes; note that, the graph $F[\pi^*]$ on $n=2k+1$ nodes can encode a watermarking integer $w$ in the range $[0, 2^{k-1}-1]$; see, authors' work \cite{CN10,CN12}.

Furthermore, in our model the code which is associated with the watermark is composed both by new code and host code; this enable us to obtain high stealth watermarked programs $P^*$. Moreover, since the watermark code has become an indispensable piece of the functionality of program $P^*$, a malicious user would need to fully understand the operations of $P^*$ in order to intervene changing possible execution flows. On the other hand, the extraction of our watermark takes into account and uses the traces of all the functions that are assigned to the nodes of the water-graph $F[\pi^*]$ which, in turn, means that if a subset of these functions is intercepted then the watermark can not be extracted; unfortunately, this implies a poor part protection of our watermarked program $P^*$.

Finally, the experimental results show the high functionality of all the Java programs $P^*$ watermarked under both the naive and stealthy cases, and also their low time complexity. The experiments also show that the watermarking approaches supported by our model can help develop efficient watermarked Java programs with respect to various and broadly used performance and resilience watermarking criteria.

Closing, we note that in light of our dynamic watermarking model WaterRPG it would be very interesting to compare it with other dynamic, or even static, already proposed software watermarking models \cite{CCDHKLS2004,CHCTS09,MIMIT00,NT04,SHKQ99}; we leave it as a direction for future work.

\bibliographystyle{abbrv}
{\small
\bibliography{WaterRPG-CCN-arXiv}}

\begin{thebibliography}{10}

\bibitem{A02}
G.~Arboit.
\newblock A method for watermarking java programs via opaque predicates.
\newblock In {\em Proc. 5th International Conference on Electronic Commerce
  Research (ICECR-5)}, 2002.

\bibitem{CN13a}
I.~Chionis, M.~Chroni, and S.~Nikolopoulos.
\newblock A dynamic watermarking model for embedding reducible permutation
  graphs into software.
\newblock In {\em Proc. 10th Int'l Conference on Security and Cryptography
  (SECRYPT'13)}, pages 74--85, 2013.

\bibitem{CN13b}
I.~Chionis, M.~Chroni, and S.~Nikolopoulos.
\newblock Evaluating the waterrpg software watermarking model on java
  application programs.
\newblock In {\em Proc. 17th Panhellenic Conference on Informatics (PCI'13)},
  pages 144--151, 2013.

\bibitem{CN10}
M.~Chroni and S.~Nikolopoulos.
\newblock Encoding watermark integers as self-inverting permutations.
\newblock In {\em Proc. Int'l Conference on Computer Systems and Technologies
  (CompSysTech'10)}, volume ACM ICPS 471, pages 125--130, 2010.

\bibitem{CN12}
M.~Chroni and S.~Nikolopoulos.
\newblock An efficient graph codec system for software watermarking.
\newblock In {\em Proc. 36th IEEE Conference on Computers, Software, and
  Applications (COMPSAC'12 Workshops)}, volume IEEE Proceedings, pages
  595--600, 2012.

\bibitem{CN12c}
M.~Chroni and S.~Nikolopoulos.
\newblock An embedding graph-based model for software watermarking.
\newblock In {\em Proc. 8th Int'l Conference on Intelligent Information Hiding
  and Multimedia Signal Processing (IIH-MSP'12)}, volume IEEE Proceedings,
  pages 261--264, 2012.

\bibitem{CN12b}
M.~Chroni and S.~Nikolopoulos.
\newblock Multiple encoding of a watermark number into reducible permutation
  graphs using cotrees.
\newblock In {\em Proc. 13th Int'l Conference on Computer Systems and
  Technologies (CompSysTech'12)}, volume ACM ICPS 630, pages 118--125, 2012.

\bibitem{CCDHKLS2004}
C.~Collberg, E.~Carter, S.~Debray, A.~Huntwork, J.~Kececioglu, C.~Linn, and
  M.~Stepp.
\newblock Dynamic path-based software watermarking.
\newblock In {\em Proc. of the ACM SIGPLAN 2004 conference on Programming
  language design and implementation}, pages 107--118, 2004.

\bibitem{CCKT03}
C.~Collberg, E.~Carter, S.~Kobourov, and C.~Thomborson.
\newblock Error-correcting graphs for software watermarking.
\newblock In {\em Proc. 29th Workshop on Graphs in Computer Science (WG'03)},
  volume LNCS~2880, pages 156--167, 2003.

\bibitem{CHCTS09}
C.~Collberg, A.~Huntwork, E.~Carter, G.~Townsend, and M.~Stepp.
\newblock More on graph theoretic software watermarks: Implementation,
  analysis, and attacks.
\newblock {\em Information and Software Technology}, 51:56--67, 2009.

\bibitem{Book-CN10}
C.~Collberg and J.~Nagra.
\newblock {\em Surreptitious Software}.
\newblock Addison-Wesley, 2010.

\bibitem{CT99}
C.~Collberg and C.~Thomborson.
\newblock Software watermarking: models and dynamic embeddings.
\newblock In {\em Proc. 26th ACM SIGPLAN-SIGACT on Principles of Programming
  Languages (POPL'99)}, pages 311--324, 1999.

\bibitem{CT2002}
C.~Collberg and C.~Thomborson.
\newblock Watermarking, tamper-proofing, and obfuscation - tools for software
  protection.
\newblock {\em IEEE Trans. Software Eng}, 28:735--746, 2000.

\bibitem{Collberg11}
C.~Collberg, C.~Thomborson, J.~Horning, W.~Silbert, L.~R. Matheson, A.~Wright,
  and S.~Owicki.
\newblock Software watermarking techniques.
\newblock {\em US Patent}, 2011/0214188, 2011.

\bibitem{CC04}
P.~Cousot and R.~Cousot.
\newblock An abstract interpretation-based framework for software watermarking.
\newblock In {\em Proc. 31st ACM SIGPLAN-SIGACT Symposium on Principles of
  Programming Languages (POPL'04)}, pages 173--185, 2004.

\bibitem{CKLS96}
I.~Cox, J.~Kilian, T.~Leighton, and T.~Shamoon.
\newblock A secure, robust watermark for multimedia.
\newblock In {\em Proc. 1st Int'l Workshop on Information Hiding}, volume LNCS
  1174, pages 317--333, 1996.

\bibitem{DM96}
R.~Davidson and N.~Myhrvold.
\newblock Method and system for generating and auditing a signature for a
  computer program.
\newblock {\em US Patent}, 5.559.884, 1996.

\bibitem{GKM1982}
S.~Graham, P.~Kessler, and M.~Mckusick.
\newblock Gprof: A call graph execution profiler.
\newblock {\em SIGPLAN Not.}, 17(6):120--126, 1982.

\bibitem{HU74}
M.~Hecht and J.~Ullman.
\newblock Flow graph reducibility.
\newblock {\em Journal of the ACM}, 21:367--375, 1974.

\bibitem{Tarjan12}
W.~Horne, U.~Maheshwari, R.~Tarjan, J.~Horning, W.~Silbert, L.~R. Matheson,
  A.~Wright, and S.~Owicki.
\newblock Systems and methods for watermarking software and other media.
\newblock {\em US Patent}, 8.140.850, 2012.

\bibitem{MIMIT00}
A.~Monden, H.~Iida, K.~Matsumoto, K.~Inoue, and K.~Torii.
\newblock A practical method for watermarking java programs.
\newblock In {\em Proc. 24th Computer Software and Applications Conference
  (COMPSAC'00)}, pages 191--197, 2000.

\bibitem{MC96}
S.~Moskowitz and M.~Cooperman.
\newblock Method for stegacipher protection of computer code.
\newblock {\em US Patent}, 5.745.569, 1996.

\bibitem{MC06}
G.~Myles and C.~Collberg.
\newblock Software watermarking via opaque predicates: implementation,
  analysis, and attacks.
\newblock {\em Electronic Commerce Research}, 6:155--171, 2006.

\bibitem{NT04}
J.~Nagra and C.~Thomborson.
\newblock Threading software watermarks.
\newblock In {\em Proc. 6th Int'l Workshop on Information Hiding (IH'04)},
  volume LNCS~3200, pages 208--223, 2004.

\bibitem{QP98}
G.~Qu and M.~Potkonjak.
\newblock Analysis of watermarking techniques for graph coloring problem.
\newblock In {\em Proc. IEEE/ACM Int'l Conference on Computer-aided Design
  (ICCAD'98)}, volume ACM Press, pages 190--193, 1998.

\bibitem{Rodriguez10}
T.~Rodriguez, B.~MacIntosh, and A.~Gustafson.
\newblock Software watermarking.
\newblock {\em US Patent}, 20100095376, 2010.

\bibitem{Sharma11}
B.~Sharma, R.~Agarwal, and R.~Singh.
\newblock An efficient software watermark by equation reordering and fdos.
\newblock In {\em SocProS (2)}, volume 131, pages 735--745, 2011.

\bibitem{SHKQ99}
J.~Stern, G.~Hachez, F.~Koeune, and J.~Quisquater.
\newblock Robust object watermarking: Application to code.
\newblock In {\em Proc. 3rd Int'l Workshop on Information Hiding (IH'99)},
  volume LNCS~1768, pages 368--378, 1999.

\bibitem{VVS01}
R.~Venkatesan, V.~Vazirani, and S.~Sinha.
\newblock A graph theoretic approach to software watermarking.
\newblock In {\em Proc. 4th Int'l Workshop on Information Hiding (IH'01)},
  volume LNCS~2137, pages 157--168, 2001.

\bibitem{XN2002}
T.~Xie and D.~Notkin.
\newblock An empirical study of java dynamic call graph extractors.
\newblock In {\em University of Washington CSE Technical Report 02-12-03},
  2002.

\bibitem{ZZZ11}
X.~Zhang, Z.~Zhang, and C.~Zhang.
\newblock Spread spectrum-based fragile software watermarking.
\newblock In {\em Nano, Information Technology and Reliability (NASNIT) 15th
  North-East Asia Symposium on}, pages 150--154, 2011.

\end{thebibliography}
\vfill

\end{document}